\documentclass[structabstract]{aa}
\usepackage{graphicx}
\usepackage{natbib}
\newcommand{\lsim}{\ \raise -2.truept\hbox{\rlap{\hbox{$\sim$}}\raise
5.truept\hbox{$<$}\ }}
\newcommand{\gsim}{\ \raise -2.truept\hbox{\rlap{\hbox{$\sim$}}\raise
5.truept\hbox{$>$}\ }}
\newcommand{\nodata}{...}
\begin{document}
   \title{VLT Optical $BVR$ observations of two bright Supernova Ia
   hosts in the Virgo cluster}

   \subtitle{Surface Brightness Fluctuations analysis}

   \author{Michele Cantiello\inst{1}
          \and
          Ilaria Biscardi\inst{1,2}
      \and
      Enzo Brocato\inst{1}
      \and
      Gabriella Raimondo\inst{1}}
\institute{INAF-Osservatorio Astronomico di Teramo, via M. Maggini
  snc, I-64100, Teramo, Italy  \\ \email{cantiello@oa-teramo.inaf.it}
         \and
             Dipartimento di Fisica -- Universit\'a  di Roma Tor
          Vergata, via della Ricerca Scientifica 1, 00133 Rome, Italy\\
             }
   \date{Received ---; accepted ---}

\authorrunning{Cantiello et al.}
\titlerunning{$BVR$ SBF analysis for NGC\,4621 and NGC\,4374}

% \abstract{}{}{}{}{}
% 5 {} token are mandatory

  \abstract
% {} leave it empty if  necessary
{}
{We study the characteristics of field stars in the two bright
  ellipticals NGC\,4621 and NGC\,4374 in the Virgo cluster to derive
  distances and stellar population properties.  Moreover, since the
  target galaxies have hosted three type Ia Supernova events, we
  investigate the correlations between the SNe\,Ia properties
  and their host stellar systems.}
{Using deep imaging $BVR$ data, obtained with the FORS2 camera mounted
  at the VLT, we analyse the Surface Brightness Fluctuations (SBF)
  properties of the targets. We adopt our measurements and existing
  empirical or theoretical calibrations to estimate the distance of
  the NGC\,4621 and NGC\,4374. For stellar population analysis, we
  measured SBF amplitudes in different galaxy regions, allowing to
  study the changes in field star properties. The three band coverage
  of present data, coupled with existing SBF measures available from
  literature, provides us with the largest wavelength coverage of SBF
  magnitudes for single objects. We present a detailed comparison
  between SBF data and models to constrain the physical
  characteristics of the dominant stellar components at $i)$ various
  galactic radii, and $ii)$ in the regions where SNe\,Ia events were
  recorded.}
{Our $V$ and $R$ SBF measures provide distances in agreement with
  literature estimates. The median of our and literature SBF-based
  distances agrees with the one from non-SBF methods, indicating the
  absence of any systematic effect in the SBF technique. Comparing
  either the SBF versus integrated colour diagrams, or the SBF versus
  SBF colour diagrams, with models, we find that stellar populations
  properties do not change significantly along galactic radius, with a
  dominant population having old age and solar chemical
  composition. The galaxies appear similar in all properties analysed,
  except for $B$-band SBF. Since the SBF magnitudes in this band are
  sensitive to the properties of the hot stellar component, we
  speculate that such behaviour is a consequence of different diffuse
  hot stellar components in the galaxies. Using specific models we
  find that the presence of a percentage of hot-HB stars in old and
  metal rich stellar populations could be at the origin of the
  observed differences. We find a good uniformity in the $V$ and $R$
  SBF and integrated colours in the regions where the three SNe\,Ia,
  presenting different absolute luminosities, exploded. On the other
  hand, the $B$-band SBF signal shows intriguing differences.}
  {}

   \maketitle

%________________________________________________________________

\section{Introduction}

Our understanding of galaxies in the distant Universe relies on how
well we understand the properties of their local templates. Thus, the
study of nearby unresolved stellar populations plays a key role to
obtain a refined characterisation of stellar populations at larger
redshifts. Although different astronomical methods exist to carry out
such detailed analysis, none of them can provide by itself robust
constraints. The presence of internal uncertainties in each method, or
calibration uncertainties, as well as the effect of the
age-metallicity degeneracy that affects most of the
spectro-photometric indicators \citep{worthey94}, prevents us from
relying on one single stellar population tracer for these studies, and
pushes the community of astronomers towards the definition of
new analysis methods.

%%%%%%%%%%%%table1%%%%%%%%%%%%%%%%%%%%%%

\begin{table*}
\caption{Main parameters of the target galaxies and observations}
\centering
\begin{tabular}{l c c}
\hline\hline
				      &   NGC\,4621   &  NGC\,4374          \\
\hline
\multicolumn{3}{c}{Galaxy parameters}\\
\hline
RA(J2000)$^{\mathrm{1}}$              &  12h42m02.3s  &  12h25m03.7s	     \\
Dec(J2000)$^{\mathrm{1}}$             &  +11d38m49s   &  +12d53m13s	     \\
Galaxy Type$^{\mathrm{2}}$	       &   E5	      &  E		     \\
Morphological Type$^{\mathrm{2}}$ &  -4.8	      &  -4.3 		     \\
Absolute $B$-band magnitude$^{\mathrm{2}}$&	-20.5 &    -21.0             \\
SNe Ia events	                      & SN1939B       &  SN1957B, SN1991bg   \\
$cz^{\mathrm{1}}$ (km/s, Heliocentric)&  410$\pm$6   &  1060 $\pm$6	     \\
$E(B{-}V)^{\mathrm{3}}$               &  0.033 mag    &  0.040 mag	     \\
$(V{-}I)^{\mathrm{4}}$		      & 1.172 $\pm$ 0.018  & 1.191$\pm$0.008 \\
\hline
\multicolumn{3}{c}{Observations}\\
Filter                               & Exposure time (s) & Exposure time  (s)\\
$B$                                  &  3375             &  2250             \\                 
$V$                                  &  3375             &  2250             \\ 
$R$                                  &  2376             &  1584             \\
\hline \hline
\end{tabular}
\begin{list}{}{}
\item[$^{\mathrm{1}}$] Data retrieved from NED (http://nedwww.ipac.caltech.edu); 
$^{\mathrm{2}}$  Hyperleda (http://leda.univ-lyon1.fr);
$^{\mathrm{3}}$ \citet{sfd98};
$^{\mathrm{4}}$  \citet{tonry01}
\end{list}

\label{tab_obs}
\end{table*}
%%%%%%%%%%%%table1%%%%%%%%%%%%%%%%%%%%%%

In the last two decades, the SBF method, introduced by
\citet{ts88} to obtain distances of early-type galaxies, has proved
being not only an accurate and precise distance indicator, but
also a powerful tracer of stellar population properties
\citep[e.g.][]{jensen03,raimondo05,cantiello07a}. By definition, the
SBF magnitude corresponds to the ratio of the second to the first
moment of the stellar luminosity function in the galaxy: $\bar{m}=-2.5
\log \bar{f}$, where $\bar{f} \equiv \frac{\Sigma_i n_i
  f_i^2}{\Sigma_i n_i f_i}$, and $n_i$ is the number of stars per bin
of flux $f_i$ \citep{ts88}.  Such definition implies that: $(i)$ SBF
magnitudes are linked to the properties of the stars in the galaxy;
$(ii)$ the SBF signal is dominated by the brightest stellar component
in the galaxy, because of the dependence on the second moment of the
luminosity function; $(iii)$ since the brightest phase in a stellar
population is wavelength-dependent, SBF magnitudes in different
pass-bands are sensitive to the properties of stars in different, and
well defined, evolutionary stages
\citep{worthey93b,brocato98,cantiello03}.

Taking advantage of archival $B$, $V$, and $R$ observations taken with
the FORS2 camera of the Very Large Telescope for two bright early-type
galaxies in the Virgo cluster, NGC\,4621 and NGC\,4374, we obtain
multi-band SBF measurements of these galaxies. These measurements,
coupled with existing ground-based $I-$ and $K-$band measures, and
space-based $F850LP$ from ACS data ($z$ hereafter; for integrated
colours we will also use ACS $F475W$ data, indicated  as $g$ in
the paper), provide the largest wavelength coverage of SBF measures
for single galaxies. With such sample of measures we investigate the
properties of the two galaxies using the SBF technique under its
twofold aspects: as a distance indicator and as a stellar population
properties tracer. Furthermore, since the target galaxies have hosted
three SNe\,Ia events -- SN\,1939B in NGC\,4621 \citep{zwicky42}, and
SN\,1957B \citep[][and references therein]{bertola64} plus SN\,1991bg
\citep{kosai91} in NGC\,4374 -- we can explore the capabilities of the
SBF method to improve our knowledge on the relation between the
SNe\,Ia progenitors and the stellar population they belong to.

The paper is organised as follows: a description of the observational
data~set, the data reduction and calibration procedures, and the
procedure for SBF measurements is given in \S \ref{sec_data}. The
analysis of distances is presented in \S \ref{sec_distances}, while
the study of stellar population properties based on SBF is presented
in \S \ref{sec_ssp}. A summary and the conclusions finish the paper in
\S \ref{sec_end}.

%__________________________________________________________________
%%%%%%%%%%%%%%%%%%%%%%%%%%%%%%%%%%%%%%%%%%%%%%%%%%%%%%%%%%%%%%
%_____________________________________________________________
%Figura Galaxies & images
%-------------------------------------------------------------
   \begin{figure*}
   \centering
   \includegraphics[width=8.5cm]{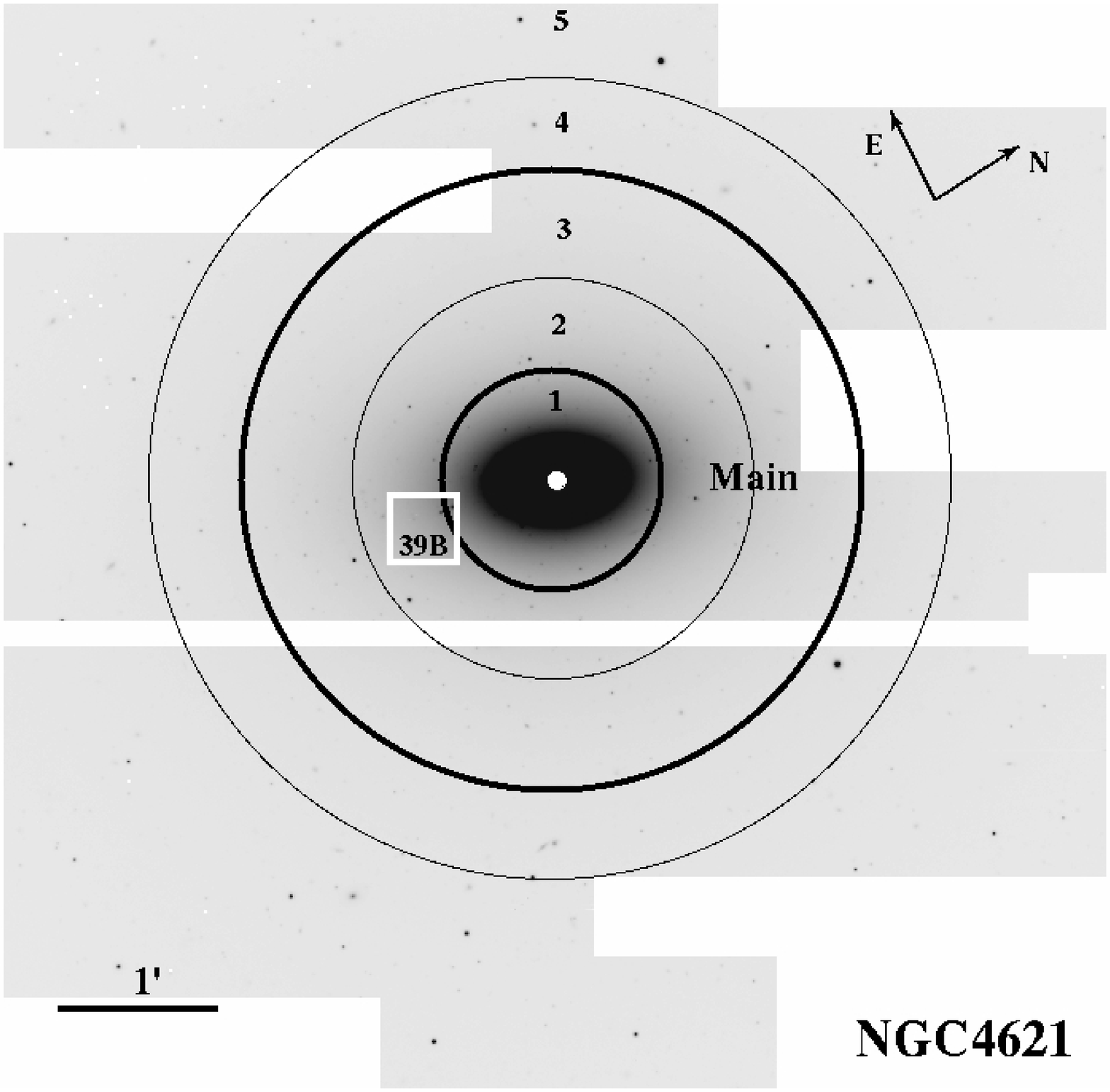}
   \includegraphics[width=8.5cm]{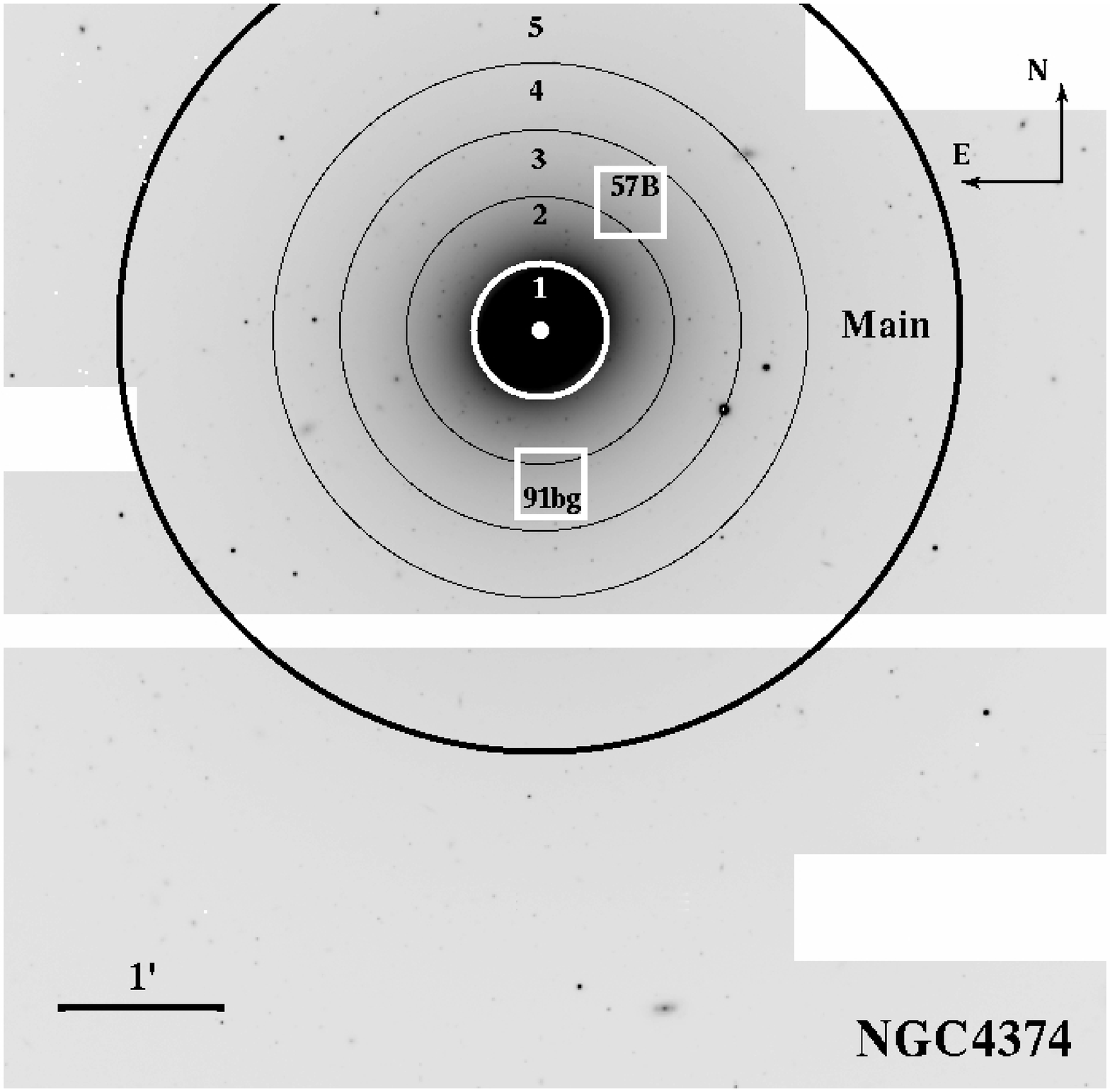}
      \caption{Left Panel: NGC\,4621 $V$-band image with the regions
        used for SBF and colour analysis. Thick-lined circles mark the
        borders of the main annulus, thin solid lines show
          the other annuli (as numbered), white boxes give the
        location of the SNe (see \S \ref{sec_sbf}). Right panel: as
        upper, but for NGC\,4374 (for sake of clarity the inner border
        of the main annulus is shown with white colour).}
         \label{images}
   \end{figure*}
%_____________________________________________________________
%Figura Galaxies & images
%-------------------------------------------------------------

\section{The Data}
\label{sec_data}
For this work we used $B$, $V$ and $R$ data of two Virgo cluster
galaxies, NGC\,4621 and NGC\,4374, observed with the FORS2 camera of
the VLT. Some relevant properties of the targets are reported in Table
\ref{tab_obs}.

The science images of the galaxies and the calibration files (bias,
flat, field of standard stars) were retrieved from the ESO
archive\footnote{http://archive.eso.org/}.

In one case, NGC\,4374, we could not use the total integration time
available because of a slight rotation ($\lsim1\deg$) between
different frames which badly affects the SBF signal (see below). In
Table \ref{tab_obs} we list the total exposure times for each filter
and for both galaxies.

\subsection{Data reduction and calibration}
\label{sec_reduc}
All data were retrieved from the archive and processed with standard
data reduction procedures using IRAF\footnote{IRAF is distributed by
  the National Optical Astronomy Observatories, which are operated by
  the Association of Universities for Research in Astronomy, Inc.,
  under cooperative agreement with the National Science Foundation.}
tasks.  The procedure is briefly described here. We obtained master
bias and flat frames (one per filter) for each observing
night. Individual frames are bias-subtracted, flattened, shifted into
registration, and combined, rejecting pixels affected by cosmic ray
hits. To combine the frames we imposed no sub-pixel registration, in
order to avoid the contamination to SBF due to the sub-pixel
interpolation  \citep[][]{jensen98}. Finally, the seeing between
targets and filters ranged between $0''.8$ and $1''.0$. The
combined frames are shown in Figure \ref{images}.

The standard calibration plan of FORS2 provides nightly multi-band
observations of \citet{landolt92} standards, which are used to
calibrate the photometry of the two galaxies.

\subsection{Data analysis and SBF measurements}
\label{sec_sbf}
To derive the photometry of sources in the frames and measure the
fluctuation amplitudes we used the same procedure described in our
previous works
\citep{cantiello05,cantiello07b,cantiello07a,cantiello09,biscardi08}.
The main steps of SBF measurement involve: sky background
determination, model and large scale residual subtraction; photometry
and masking of point-like and extended sources, including dust; power
spectrum analysis of the residual frame.

Here we briefly describe some relevant parts of the analysis.  We
determined the sky background in the galaxy images by fitting the
surface brightness profile of the galaxy with a Sersic's law 
  \citep{sersic68} plus a constant term. A first model of the galaxy
was obtained and subtracted  from the sky-subtracted frame. The
wealth of bright sources appearing after model subtraction were masked
out; the procedure of model fitting and masking was then iteratively
repeated until the residual frame (original frame minus galaxy model)
was considered satisfactory. After the subtraction of the best galaxy
model, the large scale residuals still present in the frame were
removed using the background map obtained with SExtractor
\citep{bertin96} adopting a mesh size $\sim 10$ times the FWHM
\citep{cantiello05}. In the following we refer to the sky,
galaxy-model and large scale residuals subtracted image as {\it
  residual} frame.

The photometry of fore/background sources and of Globular
  Clusters (GCs hereafter) was derived with SExtractor on the final
residual frame. As described in our previous works, we modified the
input weighting image of SExtractor by adding the galaxy model
\citep[times a factor between 0.5 and 10, depending on the galaxy; for
  details see][]{jordan04,cantiello05} so that the SBFs were not
detected as real objects. To correct for Galactic reddening we used
the $E(B{-}V)$ values from \citet{sfd98} reported in Table
\ref{tab_obs}. The aperture correction (a.c.)  was obtained from a
number of isolated point-source candidates in the frames and by making
a curve growth analysis out to $6.0''$.

%_____________________________________________________________
%Figura LLFF
%-------------------------------------------------------------
   \begin{figure}
   \centering
   \includegraphics[width=9cm]{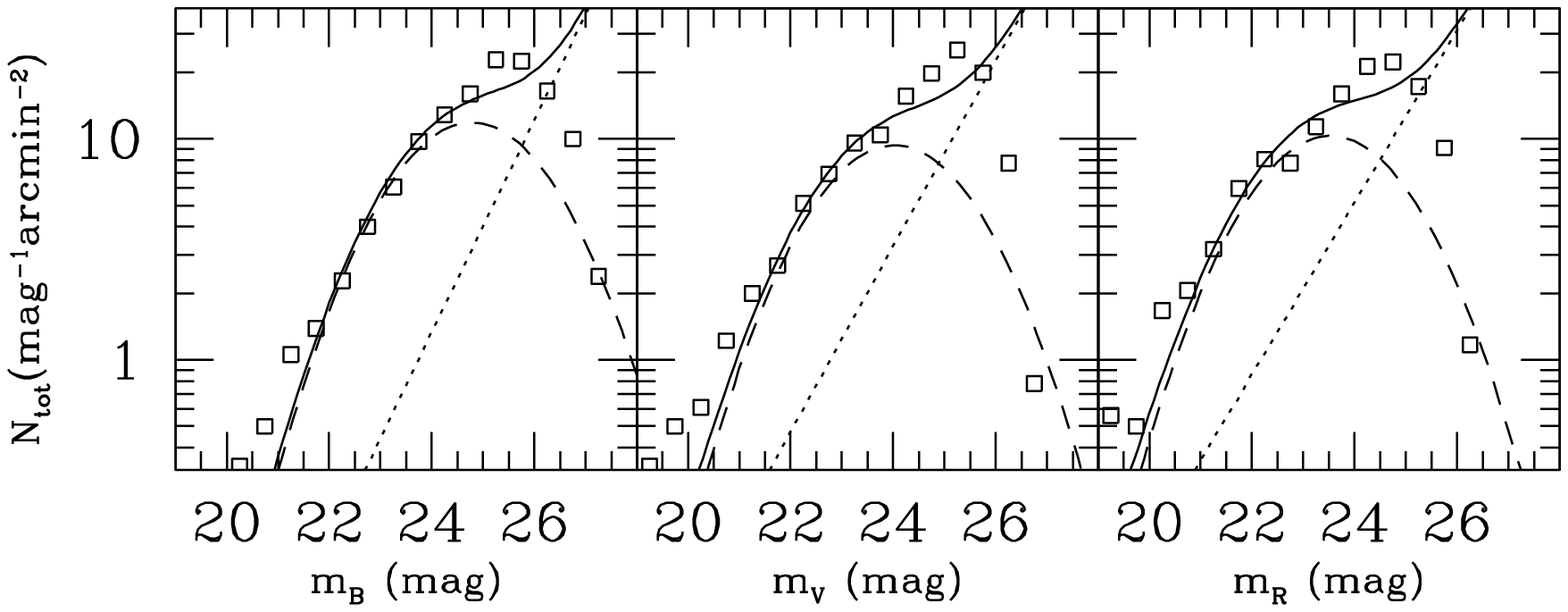}  \\
   \includegraphics[width=9cm]{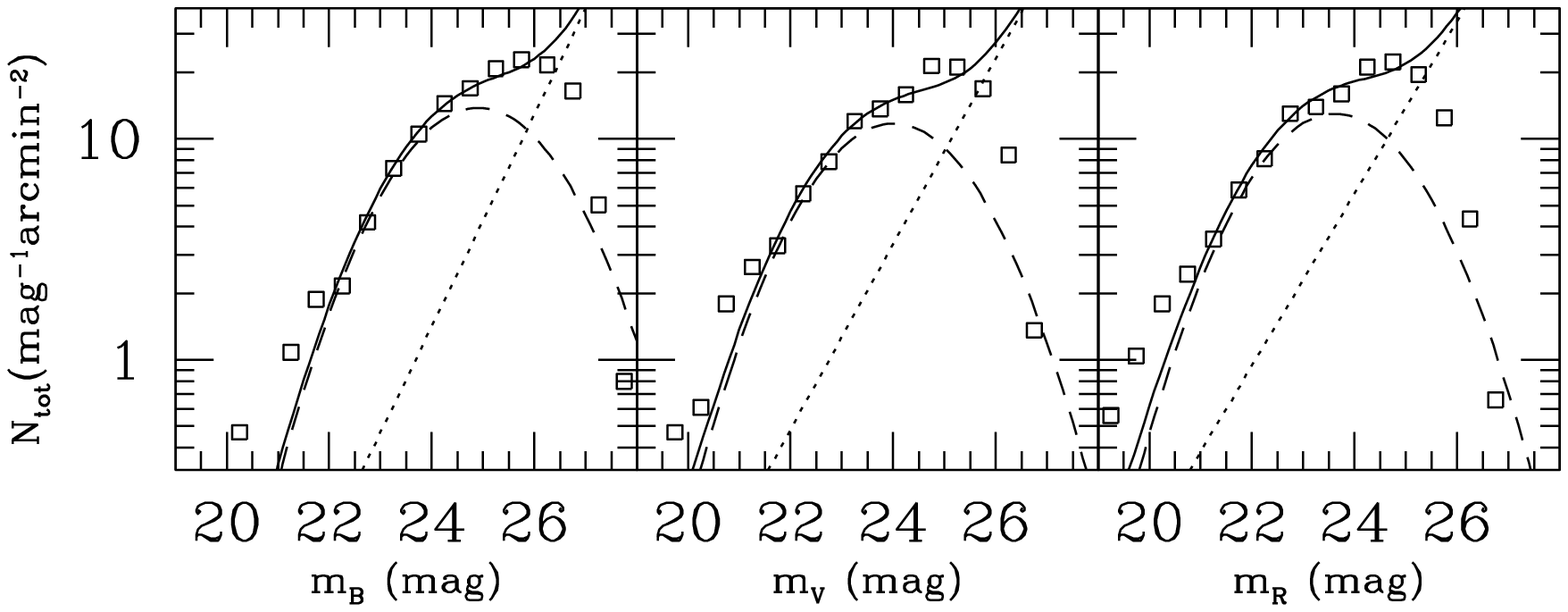}  \\
      \caption{$B$, $V$, and $R$ luminosity functions of external
      sources in NGC\,4621 (upper panels), and NGC\,4374 (lower
      panels). Open squares mark observational data; the solid lines
      represent the best fit to data. The luminosity functions of
      background galaxies and GCs are plotted as dotted and dashed
      lines, respectively.}
         \label{llff}
   \end{figure}

Once the corrected catalogue of sources was derived, the next step was
to fit the luminosity functions of external sources, that will be used
to estimate the extra-fluctuation term due to unmasked faint sources
\citep{tal90}. We derived the fit to GCs and background galaxies
luminosity functions from the photometric catalogue of sources, after
removing the brightest/saturated point-like sources and the brightest
and most extended objects. The best fit of the sum of the two
luminosity functions, shown in Figure \ref{llff}, and the
extra-fluctuation correction term, $P_r$, were derived as in
\citet{cantiello05}.

To measure SBF magnitudes we proceeded by estimating the azimuthal
average of the residual frame power spectrum, $P(k)$, then matching it
with the power spectrum of a template PSF convolved with the mask
image, $E(k)$. The total fluctuation amplitude $P_0$ was obtained via
a robust minimisation method \citep{press92} as the multiplicative
factor required to obtain the matching $P(k)=P_0 \times E(k) + P_1$,
where $P_1$ is the white noise constant term. As template PSF we used
6 different isolated bright point-like sources in each residual frame,
which, after normalisation, were singularly adopted to estimate the
SBF signal of the galaxy.

%_____________________________________________________________
%Figura POFIT
%-------------------------------------------------------------
   \begin{figure*}
   \centering
   \includegraphics[width=8cm]{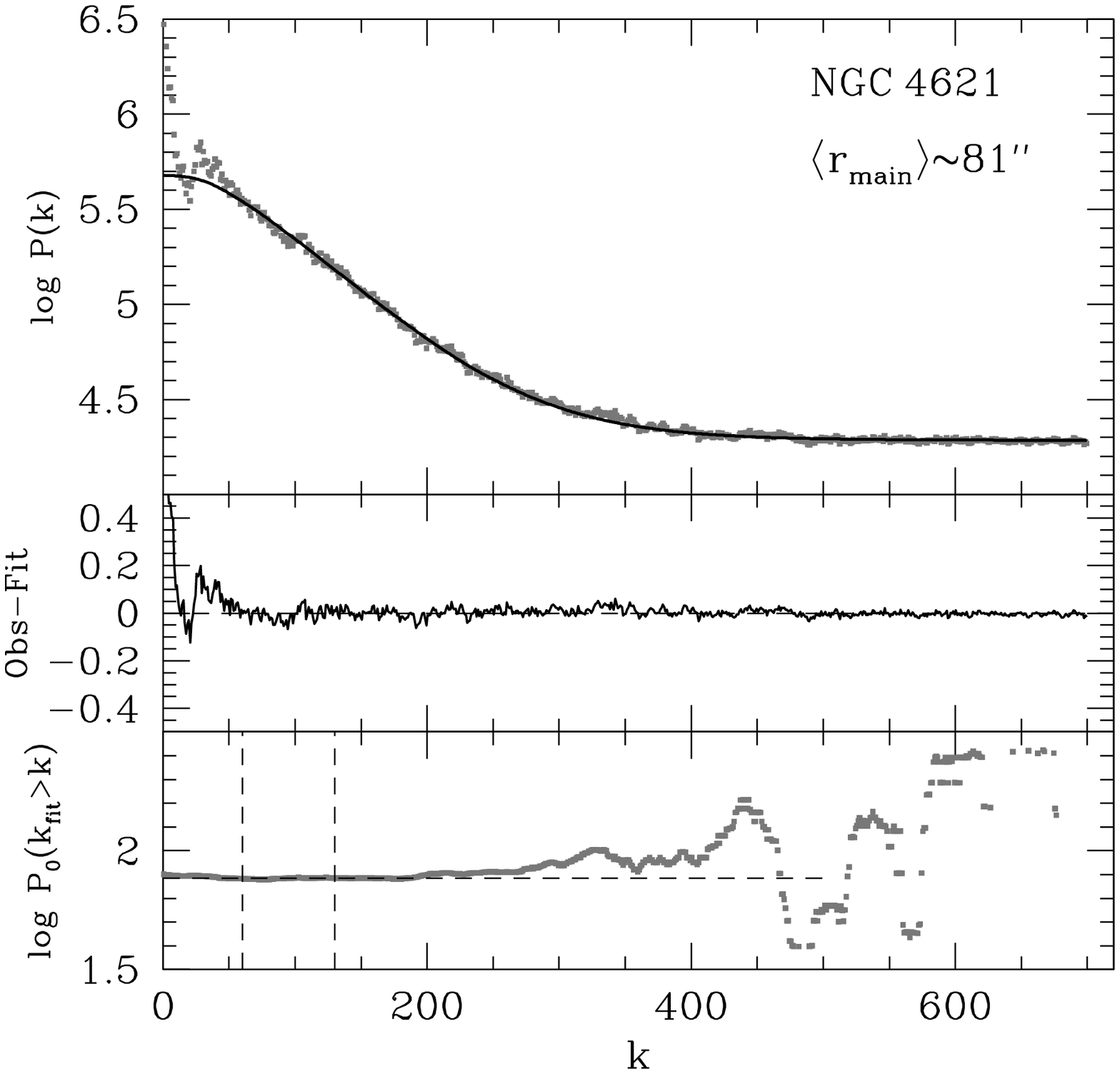}
   \includegraphics[width=8cm]{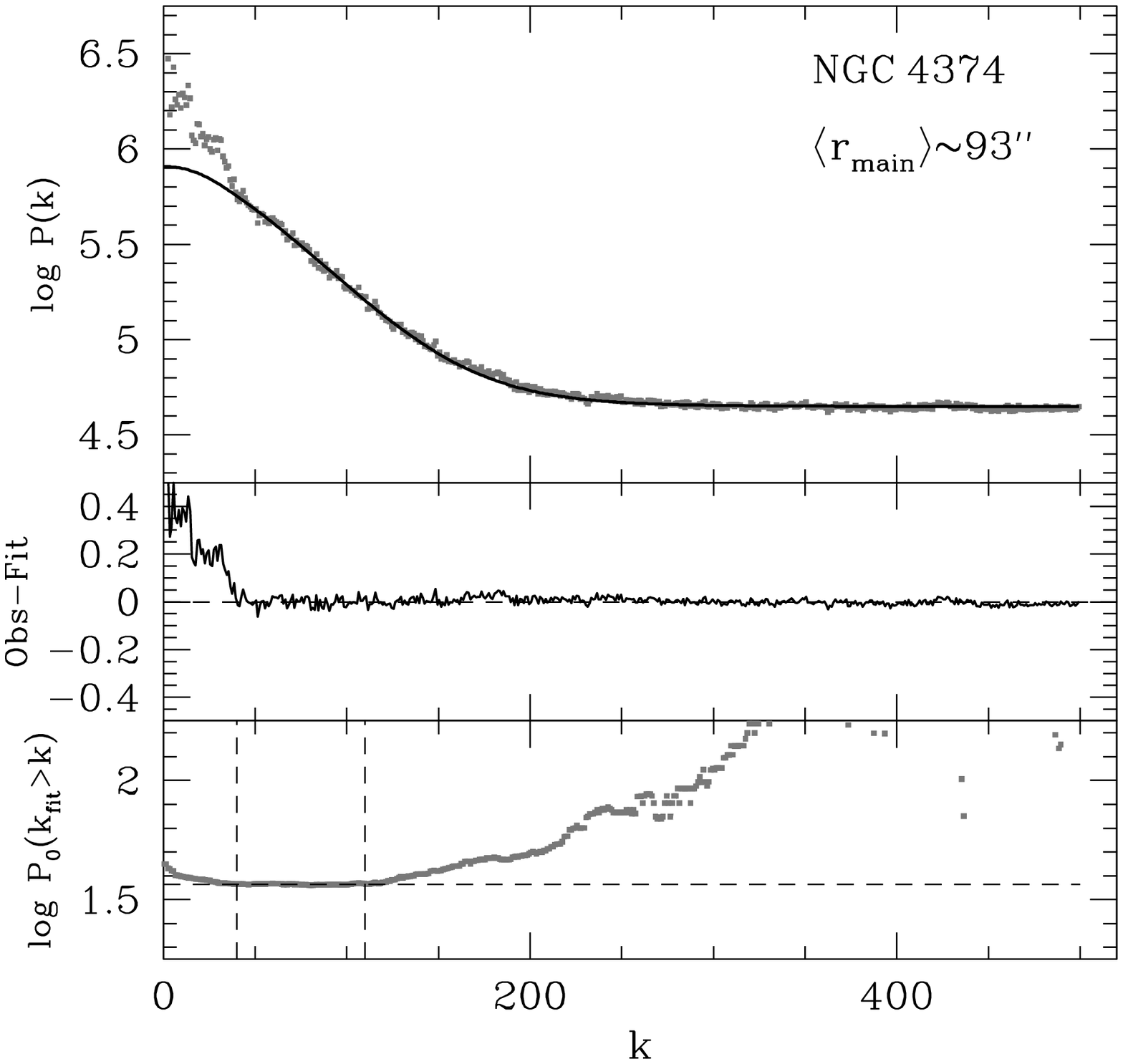}
      \caption{Power spectrum analysis of the main annuli for both
      targets. Left panels: NGC\,4621 analysis. The upper panel shows
      the logarithm of the power spectrum of the residual frame (grey
      dots) and the best fit $P(k)$ curve. In middle panel the
      difference between observed and fitted data is shown. The flat
      region of $\log~P_0(k_{fit}>k)$ between vertical dashed lines in
      the lower panel is used to evaluate the best fit parameters
      $P_0$, and $P_1$ \citep{cantiello05,biscardi08}. Right panels:
      as left, but for NGC\,4374.}
         \label{pofit}
   \end{figure*}

For both objects the fluctuation amplitude, $P_f=P_0-P_r$, was
estimated within a circular annulus, adopting the same average inner
and outer radii used by \citet{tonry01} in order to allow a
homogeneous combination of our and \citeauthor{tonry01} data. Figure
\ref{pofit} shows the power spectrum analysis of both galaxies.

Furthermore, in order to study the radial variation of SBFs, and the
SBF properties in the sites where Type Ia Supernova events are
recorded, we ran the same SBF measurement procedure described above in
five annuli per galaxy, and in three more box-shaped regions
  centred on each SN. All regions used are shown in Figure
\ref{images}.  We must emphasise that the effect of the
  extra-fluctuation correction term, $P_r$, and its relative amplitude
  with respect to the total fluctuation signal, $P_0$, changes from
  region to region, and the $P_r/P_0$ ratio increases for outer
  regions and bluer bands. Thus, a systematic under/over-estimate of
  $P_r$ may lead to over/under-estimated SBF magnitudes. Table
  \ref{tab_popr} shows the $\langle P_r/P_0 \rangle\times 100$ ratios
  for all regions and bands considered here; the numbers in the table
  demonstrate that $V$ and $R$-band SBF magnitudes have $P_r$
  corrections much lower than the total fluctuation signal, even for
  the outermost annulus considered. On the contrary, for $B$-band
  measures the $P_r/P_0$ ratio in some of the regions considered
  reaches values as high as $\sim0.45$, implying that the extra
  correction term is comparable with $P_f$, thus the SBF magnitudes
  will be considered with caution in such regions.

%%%%%%%%%%%%table2%%%%%%%%%%%%%%%%%%%%%%

\begin{table*}
\caption{$\langle P_r/P_0 \rangle$ ratios (in \%)}
\label{tab_popr}
\centering
\begin{tabular}{c |c c c | c c c}
\hline\hline 
          &   $B$   &   $V$   &   $R$    &   $B$   &   $V$   &   $R$    \\
\hline
\multicolumn{1}{c|}{Annulus} &\multicolumn{3}{c|}{NGC\,4621} &      \multicolumn{3}{c}{NGC\,4374} \\
Main	  &  17     &    7    &  5  &  29 & 11 & 9 \\  
\hline 			                               
1	  &  6      &    4    &  4  &  11 & 7  & 8 \\  
2	  &  13     &    6    &  4  &  17 & 9  & 8 \\   
3	  &  24     &    9    &  6  &  26 & 11 & 9 \\  
4	  &  34     &    13   &  8  &  38 & 13 & 9 \\   
5	  &  37     &	 15   & 10  &  44 & 16 & 11\\ 
\hline 			       
SN\,1939B  &  12    &    5    &  4  &     &    &   \\
\hline 			       
SN\,1991bg &        &         &     & 37  & 10 & 9 \\
SN\,1957B  &        &         &     & 25  &  9 & 7 \\
\end{tabular}
\end{table*}

Table \ref{tab_sbf} summarises our measurements; for each galaxy, we
tabulate (1) region identification; (2) average equivalent radius of
the region $r(arcsec)=\sqrt{(a \cdot b)}$; (3-5) SBF magnitudes
measured in $BVR$; (6-7) integrated colours.

%%%%%%%%%%%%table3%%%%%%%%%%%%%%%%%%%%%%

\begin{table*}
\caption{Surface Brightness Fluctuation and colour measurements corrected for galactic extinction.}
\label{tab_sbf}
\centering
\begin{tabular}{l c c c c c c}
\hline\hline 
Annulus & $\langle r \rangle$ &   $\bar{B}$ &  $\bar{V}$ &$\bar{R}$ & $B{-}V$ & $B{-}R$   \\
        &     (arcsec)        &    (mag)    &   (mag)    & (mag)    &  (mag)  &  (mag)    \\
\hline\hline  
\multicolumn{7}{c}{NGC\,4621} \\
Main	  &  81	    &    32.83  $\pm$ 0.10   &31.84 $\pm$ 0.06 &30.94 $\pm$ 0.08 &0.893 $\pm$ 0.002 &1.449 $\pm$ 0.002  \\
\hline 											     	          
1	  &  29     &    32.43  $\pm$ 0.09   &31.75 $\pm$ 0.07 &30.96 $\pm$ 0.06 &0.912 $\pm$ 0.001 &1.474 $\pm$ 0.001  \\
2	  &  58     &    32.83  $\pm$ 0.10   &31.83 $\pm$ 0.06 &30.94 $\pm$ 0.08 &0.900 $\pm$ 0.001 &1.455 $\pm$ 0.001  \\
3	  &  95     &    32.83  $\pm$ 0.10   &31.81 $\pm$ 0.06 &30.93 $\pm$ 0.07 &0.884 $\pm$ 0.002 &1.439 $\pm$ 0.003  \\
4	  &  132    &    32.87  $\pm$ 0.11   &31.74 $\pm$ 0.07 &30.90 $\pm$ 0.08 &0.871 $\pm$ 0.005 &1.431 $\pm$ 0.006  \\
5	  &  193    &	 32.63  $\pm$ 0.12   &31.59 $\pm$ 0.06 &30.81 $\pm$ 0.06 &0.839 $\pm$ 0.008 &1.391 $\pm$ 0.011  \\
\hline 											     	          
SN 1939B  &  55	    &    33.01  $\pm$ 0.10   &31.84 $\pm$ 0.07 &30.97 $\pm$ 0.08 &0.915 $\pm$ 0.001 &1.469 $\pm$ 0.001  \\
\hline\hline 
\multicolumn{7}{c}{NGC\,4374} \\
Main	 & 93	    &  33.31  $\pm$0.08   &  32.11 $\pm$  0.04   &31.40 $\pm$0.08   &0.925 $\pm$ 0.002 & 1.494 $\pm$ 0.003 \\
\hline 											        	        
1	 &20  	    &  32.43  $\pm$0.06   &  31.51 $\pm$  0.04   &31.01 $\pm$0.08   &0.945 $\pm$ 0.001 & 1.521 $\pm$ 0.001 \\
2	 &39  	    &  33.08  $\pm$0.07   &  32.03 $\pm$  0.04   &31.33 $\pm$0.08   &0.931 $\pm$ 0.001 & 1.501 $\pm$ 0.002 \\
3	 &63  	    &  33.36  $\pm$0.09   &  32.23 $\pm$  0.03   &31.45 $\pm$0.08   &0.922 $\pm$ 0.002 & 1.489 $\pm$ 0.004 \\
4	 &88  	    &  33.58  $\pm$0.08   &  32.23 $\pm$  0.02   &31.41 $\pm$0.08   &0.919 $\pm$ 0.003 & 1.486 $\pm$ 0.006 \\
5	 &122	    &  33.47  $\pm$0.10   &  32.22 $\pm$  0.03   &31.44 $\pm$0.08   &0.919 $\pm$ 0.006 & 1.486 $\pm$ 0.011 \\
\hline 											        	        
SN 1991bg& 58	    &  33.87  $\pm$0.07   &  32.13 $\pm$  0.04   &31.43 $\pm$0.08   &0.927 $\pm$ 0.002 & 1.495 $\pm$  0.00 \\
SN 1957B & 59	    &  33.50  $\pm$0.09   &  32.13 $\pm$  0.03   &31.35 $\pm$0.08   &0.917 $\pm$ 0.002 & 1.484 $\pm$  0.00 \\
\hline\hline 
\end{tabular}
\end{table*}

%______________________________________________________________

\section{Analysis of distances}
\label{sec_distances}
To estimate distances based on the SBF method, as for most distance
indicators, a calibration of the absolute SBF magnitude is
required. The most widely adopted bands for SBF measures are the
optical $I$ and $z$ bands, and the near--$IR$ $H$ and $K$ ones
\citep{pahre94,tonry01,mei07vii,jensen03,blakeslee10b}. All $z$ and
$H$ measures available come from HST observations taken with the
ACS/WFC and NICMOS/NIC2 detectors, respectively
\citep[e.g.][]{jensen03,mei07vii,blake09}. The total number of SBF
data available in these pass-bands is around $\sim450$ measures. As a
consequence, the calibrations in such bands, especially the optical
ones, are well established as testified, for example, by the accurate
characterisation of the $\bar{z}$ versus $(g{-}z)$ relation derived by
\citet{blake09} using ACS data of Virgo and Fornax cluster galaxies.

SBF measurements in other bands are not {\it popular}, with a total of
$\sim$70 measures available for $B$, $V$ and $R$ bands
\citep[e.g.][]{tal90,jerjen00,bva01,sodemann96,cantiello07a}. This is
due to the fact that SBF magnitudes at these wavelengths tend to be
more sensitive to stellar population properties and, consequently, the
calibration has a larger scatter and is less reliable, while for
distance studies the most favourable condition is that the magnitude
of the standard candle is relatively constant or has a tight
  predictive relation as a function of some other
  distance$-$independent property, such as colour. Taking advantage
of the well constrained distances of the target galaxies, we will
derive new distances based on our measurements and
empirical/theoretical calibrations, and compare the various results to
test calibrations adopted. In the upper part of Table
\ref{tab_distances} we report the SBF measures for NGC\,4621 and
NGC\,4374 in different bands as derived in literature, the distance
moduli $m-M$, and the calibration used. The second part of the table
gives the distances derived by us as discussed below.

\subsection{Empirical calibrations}
\label{emprical}
\citet{cantiello07b} compared the existing $I$ and $V$ band
calibration equations derived from different observational data, in
order to identify the best empirical calibration for the colour
interval $0.85 \leq V{-}I \leq 1.30$ mag in both pass-bands. For $V$-band
SBF magnitudes the best calibration equations resulted to be:
\\
$\bar{M}_V=0.81 \pm 0.12 + (5.3 \pm 0.8) \cdot [(V{-}I)-1.15]$ for $1.05 \leq V{-}I \leq 1.30$ mag (1);
\\
and\\
$\bar{M}_V=-0.50 \pm 0.27$ mag for the  $0.85 \leq V{-}I
\leq 1.05$ mag (2).

To use the above calibration we adopt the $V{-}I$ colours from
\citet{tonry01}, and the SBF measures listed in Table \ref{tab_sbf}
for the main annulus. The distances obtained are shown in Table
\ref{tab_distances} and will be discussed with all other estimates at
the end of the section.

\smallskip

For $R$-band SBF, the most recent empirical calibration for
  normal elliptical galaxies was provided by \citet{tal90} who,
however, considered it unreliable. Distance estimates for a large
number of dwarf ellipticals have been provided by Jerjen and
collaborators \citep[e.g.][]{jerjen98,jerjen00,jerjen04} based on
$R$-band SBF measures and semi-empirical relations. Using the $B{-}R$
colour, the authors recognised two different branches for the SBF
versus colour relation, a linear branch: \\ $\bar{M}_R=- 8.94 + 6.09
\cdot (B{-}R)$ for $1.10 \leq B{-}R \leq 1.50$ mag (3); \\ and an
parabolic branch, partly overlapping to the previous one:
\\ $\bar{M}_R=-1.39 + 1.89 \cdot [(B{-}R)-0.77]^2$ for $0.80 \leq
B{-}R \leq 1.35$ mag (4) \citep{jerjen04}.  To derive distances,
reported in Table \ref{tab_distances}, we used the red/linear branch
calibration.

\smallskip

The small amplitude of the fluctuations in the $B$-band and the higher
sensitivity to stellar population properties with respect to other
optical bands \citep[see,
  e.g.,][]{worthey93b,cantiello03,cantiello07a} make this band 
  unreliable for distances studies. Besides, only a handful of
$\bar{B}$ measures exist, including the present ones, and no distance
has been estimated using $B$-band SBF measurements.

\subsection{Theoretical calibrations}
\label{theory}
Empirical calibrations are generally preferred for SBF studies, also
thanks to the aforementioned small uncertainties for certain filters.
However, the derivation of such relations requires substantial
  observational effort for each pass-band. On the other hand, models
have the advantage of being homogeneous through the different bands,
but need many different counter-checks to be considered reliable.  In
this work we take as reference the SBF versus colour equations derived
using the simple stellar population (SSP) models from the
Teramo-Stellar Population Tools (SPoT group\footnote{Visit the
  web-site: www.oa-teramo.inaf.it/spot}. For a detailed review of the
SPoT models we remind the reader to \citet{raimondo05} and references
therein.  These models have already been proved being very effective
not only to match the empirical SBF calibrations in different bands,
but also to reproduce the resolved (colour magnitude diagrams) and
unresolved (colours, magnitudes) properties of stellar populations
\citep{brocato00,cantiello03,raimondo05,cantiello07b,biscardi08,cantiello09,raimondo09}.

For example, in a previous paper, \citet{biscardi08} showed that
the SPoT models for an age interval between 1.5 and 14 Gyr, and
metallicity $[Fe/H]$ between -0.4 and 0.3 dex are able to reproduce
both the $I$-band and $z$-band calibrations from \citet{tonry01} and
\citet{mei07vii}, respectively.

In the present work we adopt the same models used by
\citeauthor{biscardi08} and derive absolute $V$ and $R$-band SBF
magnitudes versus the integrated $B{-}R$ colour.  Using the on-line
SPoT models we obtain the following relations:
\\ $\bar{M}_R=-4.11+2.84\cdot(B{-}R)$ (5), \\ $\bar{M}_V=-2.69 +
2.60\cdot(B{-}R)$ (6), \\ we also derive, for sake of completeness,
the $B$-band calibration: \\ $\bar{M}_B=-0.68 + 2.32\cdot(B{-}R)$ (7).
However, we recall that it is hazardous to derive distances from
eq. (7) because $\bar{M}_B$ strongly depends on the properties of the
stellar population originating the SBF signal (see section
\S\ \ref{sec_ssp_hot}).

As a further check based on an independent set of stellar population
models, we also used the $V$ and $R$ equations provided by
\citet{bva01}, which, differently from the SPoT models, are derived
using composite stellar populations (see the quoted paper for more
details).

The distance moduli of the two targets, obtained using both the SPoT
and the \citet{bva01} theoretical calibrations, are reported in Table
\ref{tab_distances}.

\subsection{Results}
\label{results}
By inspecting the new and old distance estimates in Table
\ref{tab_distances} we find a satisfactory agreement within the quoted
uncertainties, no matter what calibration (empirical or theoretical)
or pass-band/colour relation is used, with the sole exception of
$B$-band data which we report here only for sake of completeness.

The distance moduli derived from SBF are affected by the
uncertainties present in the empirical/theoretical calibrations,
besides the uncertainty of the SBF measure itself. In this work, for
the distance moduli taken from literature where no calibration error
is given we consider safe to assume an error of the order of
$\sim$0.20 mag, which includes zero-point uncertainty and the scatter
of empirical calibrations \citep{tonry01,jensen03}. Similarly we
assume $\sim$0.20 mag error for the theoretical calibrations, which is
originated by the spread between models with different age and
metallicity.

Keeping in mind such uncertainty, and the error in the SBF measures,
the results in Table \ref{tab_distances} can be summarised as follows:
\begin{itemize}

\item although the colour provided by \citeauthor{tonry01} was
  measured in galaxy regions slightly different from ours\footnote{As
    explained before, we adopted the same ``average'' inner and outer
    radii of \citeauthor{tonry01}, but the detailed shape of the
    annulus, plus the masking of sources is clearly different between
    the two datasets.}  the distances derived from $\bar{V}$ using
  eq. (1) agree with other data from literature. A similar result is
  true for $\bar{R}$ if the linear branch relations by
    \citet{jerjen04} are adopted;

\item the distance estimates obtained using empirical calibrations
  show a slight tendency to have a smaller scatter with respect to
  those from theoretical calibrations;

\item whether SSP models from the SPoT group or the \citet{bva01}
  composite models are used the distances derived are substantially
  similar to each other, and agree with expected values. Such result
  implies that for the bandpasses and the colour interval
    considered here composite stellar population models, which try to
  better reflect the real population mixing of galaxies, are not
  strictly necessary for the purposes of deriving appropriate SBF
  versus colour relations;

\item as discussed above, coupling $\bar{B}$ measures with eq. (7)
  provides unreliable distance moduli, this confirms that this band
  must discarded for distances. Interestingly, the difference
  between the $B$-band distance moduli and the literature average
  moduli is significantly larger for NGC\,4621 ($\Delta(m-M)=
  (m-M)_{B,SBF}-\langle m-M \rangle_{lit.}\sim -0.8$ mag) than for
  NGC\,4374 ($\Delta (m-M)\sim -0.6$ mag). We will analyse further
  this evidence in the next section on stellar population analysis.

\end{itemize}

Finally, the last lines of Table \ref{tab_distances} provide the
median distance moduli of the two targets obtained by averaging $i)$
our estimates (empirical and theoretical calibrations, except
$B$-band), $ii)$ all SBF $m-M$, including ours, $iii)$ the 
  distance moduli without SBF\footnote{All non-SBF distances are
    taken from the NED Redshift Independent Distance database. The
    distance indicators used include Globular Cluster and Planetary
    Nebulae Luminosity Functions, Type Ia Supernovae and Globular
    Cluster half light radii. For a complete list of methods, and
    associated references, visit the URL
    http://ned.ipac.caltech.edu/forms/d.html. In case of multiple
    estimates obtained with the same indicator only the most recent
    one is considered.}, and $iv)$ the results from ours plus all
literature data. By inspecting such data we find an excellent
  agreement between our and literature distance moduli and, more in
  general, between non-SBF and SBF-based distances, a results that
  should be regarded as a direct evidence of the absence of any
  significant bias or systematics between the quoted distances. In
  addition, such agreement also proves the reliability of $V$-and
  $R$-band SBF calibrations presented in this section.

%%%%%%%%%%%%table4%%%%%%%%%%%%%%%%%%%%%%

\begin{table*}
\caption{Galaxy distances from this work and from literature.}
\begin{tabular}{c c c | c c | c c}
\hline \hline 
\multicolumn{3}{c}{}                       &\multicolumn{2}{|c|}{NGC\,4621 }  & \multicolumn{2}{c}{NGC\,4374}  \\
Filter  & Colour       &  Ref. Color/Data & $\bar{m}$ & $m-M$                               & $\bar{m}$ & $m-M$             \\
\hline\hline
\multicolumn{7}{c}{SBF from literature}\\
\hline
$I$     & $V{-}I $     & [1]              & 29.67$\pm$  0.18             &31.16$\pm$0.20  &29.77$\pm$             0.09 &31.16$\pm$0.11  \\
$z$     & $g{-}z $     & [2]              & 29.12$^{\mathrm{a}}\pm$  0.01 &30.86$\pm$0.07  &29.53$^{\mathrm{a}}\pm$  0.01 &31.34$\pm$0.07  \\
$K$     & \nodata      & [3]              & 25.46$\pm$  0.16             &\nodata         &25.43$\pm$  0.22            &\nodata         \\
\hline \hline 
\multicolumn{7}{c}{Our measurements }\\
\hline 
\multicolumn{7}{c}{Empirical Calibrations$^{\mathrm{b}}$ }\\
\hline
$V$ & $V{-}I$ &  [4]   &31.84$\pm$ 0.06 & 30.91$\pm$0.21 &  32.11$\pm$ 0.04  &31.08$\pm$0.20  \\
$R$ & $B{-}R$ &  [5]   &30.94$\pm$ 0.08 & 31.05$\pm$0.21 &  31.40$\pm$ 0.08  &31.25$\pm$0.22  \\
\hline	   
\multicolumn{7}{c}{Theoretical Calibrations--SSP$^{\mathrm{b}}$}\\
\hline
$B^{\mathrm{c}}$ & $B{-}R$  & [6]& 32.83 $\pm$0.10   &30.2  & 33.31 $\pm$0.08  &30.5 \\	
$V$ & $B{-}R$  & [6]& 31.84 $\pm$0.06   &30.76$\pm$0.21 & 32.11 $\pm$0.04  &30.91$\pm$0.20 \\
$R$ & $B{-}R$  & [6]& 30.94 $\pm$0.08   &30.93$\pm$0.21 & 31.40 $\pm$0.08  &31.27$\pm$0.22\\
\hline
\multicolumn{7}{c}{Theoretical Calibrations--CSP$^{\mathrm{b}}$}\\
\hline
$V$ & $V{-}I$  & [7]  & 31.84$\pm$ 0.06   &30.99$\pm$0.21 & 32.11$\pm$ 0.04  &31.17$\pm$0.20    \\
$R$ & $V{-}I$  & [7]  & 30.94$\pm$ 0.08   &31.04$\pm$0.21 & 31.40$\pm$ 0.08  &31.41$\pm$0.22    \\
\hline \hline
\multicolumn{7}{c}{Median $m-M$}\\
\hline
\multicolumn{3}{c}{}                       &\multicolumn{2}{c}{NGC\,4621 }  & \multicolumn{2}{c}{NGC\,4374}  \\
\multicolumn{3}{c}{This work}              &\multicolumn{2}{c}{30.96 $\pm$0.11} & \multicolumn{2}{c}{31.21$\pm$ 0.17} \\  
\multicolumn{3}{c}{all SBFs}               &\multicolumn{2}{c}{31.08 $\pm$0.19} & \multicolumn{2}{c}{31.21$\pm$ 0.14} \\ 
\multicolumn{3}{c}{No SBFs }               &\multicolumn{2}{c}{30.97 $\pm$0.21} & \multicolumn{2}{c}{31.17$\pm$ 0.18}  \\
\multicolumn{3}{c}{All avaliable $m-M$} &\multicolumn{2}{c}{30.98 $\pm$0.17} & \multicolumn{2}{c}{31.22$\pm$ 0.16}  \\

\hline \hline
\end{tabular}

\begin{list}{}{References}
\item
[[1]] \citet{tonry01};
[2] \citet{blake09};
[3] \citet{pahre94};
[4] \citet{cantiello07b};
[5] \citet{jerjen04};
[6] this work;
[7] \citet{bva01}.
\end{list}
\begin{list}{}{Notes}
\item
[$a$] \citet{blake09} data are given in the AB-mag
system.  \item [$b$] We adopted $\sim$0.2 mag as safe
calibration error for $i)$ distance moduli given in literature with no
calibration uncertainty, and $ii)$ distances based on theoretical
calibrations (see text).
 \item [$c$] Distances derived from $B$-band SBF are reported to show that the sensitivity of $\bar{B}$ to stellar population 
poperties makes unreliable any distance estimate.
\end{list}
\label{tab_distances}
\end{table*}

\subsection{Comparison with SNe\,Ia distances}

The comparison of SBF and SNe\,Ia distances plays a fundamental role
in the problem of the cosmological distance scale. SBF magnitudes are
capable of providing distances with $\lsim10\%$ accuracy within 100
Mpc \citep{jensen01,biscardi08,blakeslee10b}, a result that will be
likely improved with new and next generation observing facilities,
allowing to estimate accurate distances of bright ellipticals out to
$\sim$200 Mpc. On the other hand, SNe\,Ia can provide distances to
objects at much larger distances. Thus, deriving self-consistent
distances using these two indicators is a crucial step to bridge local
to cosmological distances, in order to reduce the number of rungs in
the cosmological distance scale, i.e. to reduce systematic
uncertainties.

The sample of objects with known SBF measures and well studied SNe\,Ia
light curves is relatively small, due to the fact that SNe\,Ia occur
in all kind of galaxies, but are mostly observed in late type galaxies
because of an observational bias, on the contrary SBF are measured
almost only in early type galaxies.

Nevertheless, a comparison of SBF and SNe\,Ia distances was carried
out by \citet{ajhar01}, who found that there is a good agreement
between the two distance indicators, provided that a consistent set of
Cepheid calibrating galaxies is used. However, their statistics
grounded on $\sim 10$ objects and \citeauthor{ajhar01} made it clear
that {\it ``[...] Unquestionably, the SN\,Ia and SBF absolute
  calibrations are in need of further refinement.''}

As already mentioned, a total of three type Ia Supernova events has
been recorded in the two galaxies.  Two SN\,Ia have been observed in
NGC\,4374, SN\,1991bg and SN\,1957B, while SN\,1939B was discovered in
NGC\,4621.  The SNe are located all at $\sim$1 arcmin from the
photometric centre of the host galaxy. Even though all three SNe are
classified as type Ia, they show a wide range of luminosity -- not
unexpected in E/S0 galaxies \citep{gallagher05} --, in particular all
events are fainter at maximum light than a normal SN\,Ia after
correcting for the correlation between peak luminosity and decline
rate \citep{hamuy96}.  SN\,1991bg is one of the most sub-luminous
SN\,Ia yet observed \citep[$\sim 2.5$ mag fainter than
  normal,][]{turatto96}. SN\,1957B has an absolute magnitude at
maximum $\sim 0.2$ mag fainter than a normal SN, but still brighter
than SN\,1991bg \citep{howell01}. Finally, SN\,1939B at maximum light
is $\sim 0.6$ mag sub-luminous \citep{minkowski64}.

Unfortunately, being sub-luminous events, the standard peak
luminosity versus decline rate relation does not provide good distance
estimates with these SNe. \citet{ajhar01}, in fact, did not take into
account SN\,1939B and SN\,1957B, while SN\,1991bg appeared in their
list of Supernovae but it was not used for SBF-SNe\,Ia comparison, it
is was instead included {\it ``for completeness and for future studies
  of SNe\,Ia luminosities.''.}

More recent studies, however, have provided new methods that can be
adopted to derive distances with these sub-luminous SNe: the
calibration based on decline rate and colours by \citet{folatelli10},
the MLCS2k2 by \citet{jha07}, and the $\Delta C_{12}$ method by
\citet{wang06}.

Using eq. (6) of \citet{folatelli10} with the best-observed fit
parameters, we obtain the following distance moduli:

\begin{itemize}

\item SN\,1939B in NGC\,4621: taking the peak luminosity from the
  Asiago Supernova Catalogue \citep{barbon89}\footnote{The updated
    catalogue is available at the URL http://graspa.oapd.inaf.it}
  $m_{B,max}=12.3$, a decline rate of $\Delta m_{15}\sim 1.75$
  estimated from the light curve reported in \citet{leibundgut91}, and
  using eq. (3) from \citet{folatelli10} to get the
  $(B^{max}-V^{max})_0$ colour, we obtain $m-M=31.0$, with an
  uncertainty of the order of $\sim$0.5 mag if errors of $\sim$0.2
  mag, 0.25 days and 0.15 mag are adopted for $m_{B,max}$, $\Delta
  m_{15}$ and $(B^{max}-V^{max})_0$, respectively;

\item SN\,1957B in NGC\,4374: adopting $m_{B,max}=12.20 \pm 0.14$
  \citep{lanoix98}, $\Delta m_{15}=1.3$ \citep{howell01}, and
  $(B^{max}-V^{max})_0=0.01$ \citep[using eq. (3) from][]{folatelli10}
  the distance modulus is $m-M=31.1$, with uncertainty $\sim$0.3 mag
  adopting an error of $\sim$0.1 on both the decline rate and on
  $(B^{max}-V^{max})_0$;

\item SN\,1991bg in NGC\,4374: with $m_{B,max}=14.75$ and
  $m_{V,max}=13.95\pm0.02$ \citep{turatto96}, $\Delta m_{15}=1.93 \pm
  0.10$ \citep{phillips99} we derive $m-M=31.1\pm0.3$;

\end{itemize}

In all cases the distance moduli derived using the calibration from
\citet{folatelli10} agree within uncertainties with the median values
reported in Table \ref{tab_distances}.

\citet{jha07} developed an updated version of the MLCS method
\citep{riess98} called MLCS2k2, which includes new procedures for the
$K$-correction and for internal extinction corrections. Using the
light curve parameters of SN\,1991bg \citeauthor{jha07} derive
$m-M=31.42\pm0.10$ adopting $H_0=65~Km~s^{-1}~Mpc^{-1}$. Although such
distance agrees with the values reported in Table \ref{tab_distances},
the agreement becomes certainly better if $H_0=73~Km~s^{-1}~Mpc^{-1}$
is taken - a value consistent with the ones typically derived from SBF
distances \citep{tonry00,freedman01} - as suggested by the authors; in
that case, in fact, one has $m-M=31.17\pm0.10$ mag.

In the case of \citet{wang06} the calibration parameter adopted is the
$\Delta C_{12}$, i.e. the $B{-}V$ colour at 12 days past optical
maximum. The authors find $m-M=30.35\pm0.19$, based on the properties
of SN\,1991bg and using $H_0=72~Km~s^{-1}~Mpc^{-1}$. Such $m-M$ is
more than 2--$\sigma$ out the median values in Table
\ref{tab_distances} obtained using many independent distance
indicators, and it also disagrees with the values obtained using the
\citet{folatelli10} or \citet{jha07} calibrations, thus we flag the
\citeauthor{wang06} estimate as unreliable for the case of SN\,1991bg.

In conclusion, we find that the distances of NGC\,4621 and NGC\,4374
derived via the light curve properties of their SN\,Ia agree perfectly
with the SBF distances -- and with the most recent estimates from
literature -- if the calibration by \citet[][useful for all three
  SNe]{folatelli10}, or \citet[][for the case of SN\,1991bg]{jha07}
are used.

%______________________________________________________________

\section{Analysis of unresolved stellar populations with SBF and
  integrated colours}
\label{sec_ssp}

As mentioned before, the SBF magnitude of a stellar population
corresponds to the ratio of the second to the first moment of its
luminosity function. Since its first applications, such characteristic
has suggested the use of the SBF method as a powerful diagnostics of
stellar population properties. The earliest theoretical studies on
this subject \citep{buzzoni93,worthey93b,worthey93a} were focused on
the relationship between SBF and age/metallicity, but also showed how
the fluctuation amplitude in bluer pass-bands, like $B$, could be used
as a useful tracer of the hot stellar components in the galaxy.
Despite several authors have confirmed, and further extended, such
early findings \citep{bva01,cantiello03,gonzalez04,raimondo09}, a
comprehensive and homogeneous analysis of multi-band SBF measurements
for a large sample of galaxies is still missing. The various set of
models have clearly shown that optical to near--$IR$ SBF colours can
be very effective to lift the age/metallicity degeneracy which affects
``classical'' integrated colours \citep{worthey94}. Observationally, a
few $\bar{I}-\bar{H}$ data exist in literature \citep{jensen03}, but
they mostly refer to spatially non homogeneous regions. On the other
hand, $\bar{V}-\bar{I}$ colours have also been studied for various
galaxies, but they suffer for an age/metallicity degeneracy
  similar to classical colours, although the information obtained
using such measures is independent and complementary to the integrated
colours derived for the same targets
\citep{tal90,bva01,cantiello07b,cantiello07a}.

Concerning the use of SBF to study stellar populations, using high
quality data has allowed to measure SBF variations in optical bands
within different regions of the galaxy. As a consequence, even if $V$
and $I$ SBF magnitudes and colours are not as much effective in
removing the age/metallicity degeneracy (see below), the measure of
SBF radial variations has provided useful results on how the mean
properties of the dominant stellar population change with galactic
radius \citep{cantiello05}.

The targets analysed in this work have already been considered for
other SBF measurements surveys. In particular, the aforementioned
ground-based SBF survey \citep{tonry01}, the near--$IR$ measurements for
nine Virgo ellipticals by \citet{pahre94}, and the ACS Virgo Cluster
Survey \citep{cote04} all have the two galaxies in their target list.
The SBF magnitudes from these databases are reported in the first part
of Table \ref{tab_distances}.  None of these cited works presents the
measurement of SBF magnitudes at various galactic radii.  On the
contrary, thanks to the high quality of the VLT data available, we
have been able to obtain fluctuation amplitudes at various radii
(Table \ref{tab_sbf}).

Differently from other bright elliptical galaxies studied in
previous works \citep{cantiello05,cantiello07b}, we can report the
absence of radial $\bar{V}$ and $\bar{R}$ gradients in 
  NGC\,4374. For NGC\,4621, instead, a small but non-negligible
  brightening of SBF magnitudes at larger radii is found in $V$ and
  $R$. On the other hand, the $\bar{B}$ of both galaxies shows a
  significant scatter between the different annuli, thus no evidence
  of systematic trends with radius is observed in this band.

%_____________________________________________________________
%Figura SBF-rad new SPoT models
%-------------------------------------------------------------
   \begin{figure*}
   \centering
   \includegraphics[width=15cm]{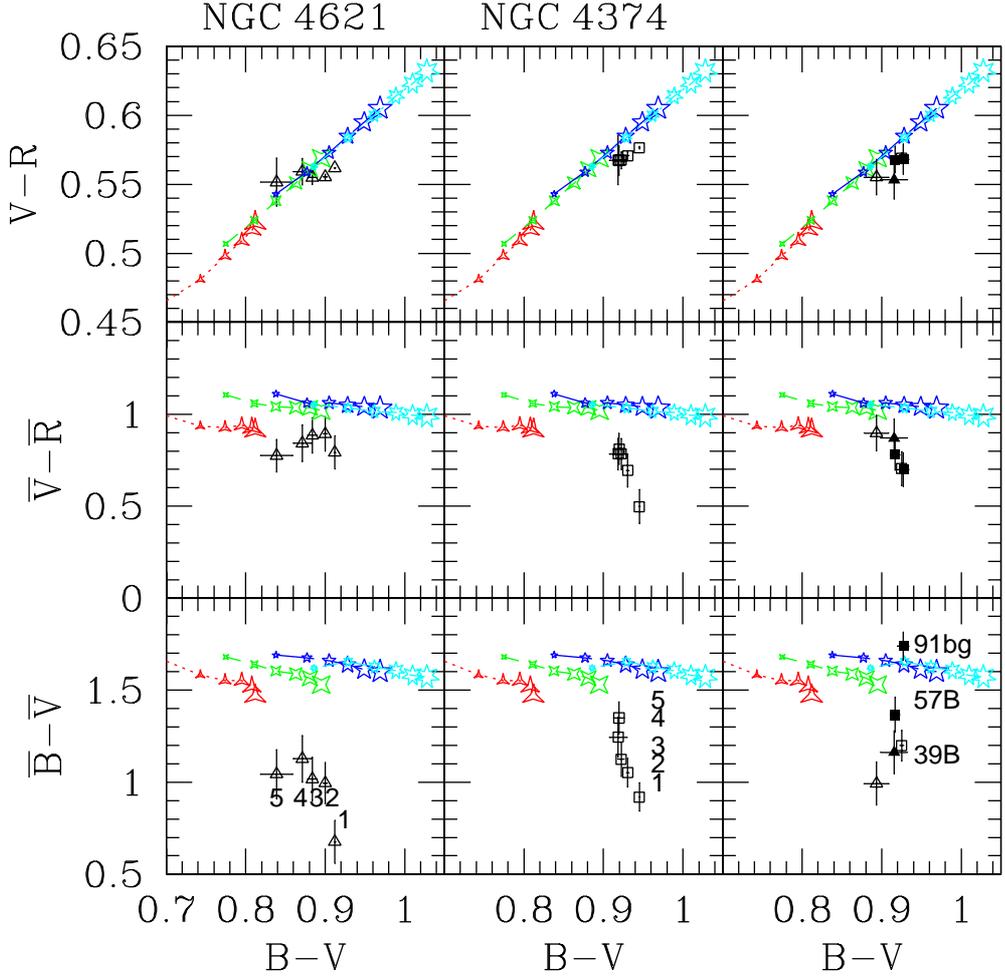}
      \caption{SBF-colours and integrated-colours measured for
      NGC\,4621 (left panels) and NGC\,4374 (middle panels), compared
      to the SPoT models from \citet{raimondo11}.  Black symbols mark
      observational data for the various annuli as labelled (triangles
      for NGC\,4621, squares for NGC\,4374). Measures within regions
      hosting SNe, and the measure for the main annulus for both
      galaxies are shown in right panels: NGC\,4621 and its SN\,Ia
      (SN\,1939B) are shown with a full and empty triangle,
      respectively; NGC\,4374 and its SNe\,Ia (SN\,1957B, and
      SN\,1991bg) are shown with full and empty squares,
      respectively. For the models, different $[Fe/H]$ are indicated
      with different symbols ($[Fe/H]$= -0.7, -0.4, 0.0,+0.4 dex with
      three, four, five and seven pointed stars, respectively). Larger
      symbols size mark older ages, from 4 to 14 Gyr with 2 Gyr
      step. \it{[See electronic version of the Journal for a colour
      version of the figure.]}}
         \label{sbf_rad}
   \end{figure*}

\subsection{SBF-colours versus integrated colours}
\label{sec_ssp_hot}

By plotting SBF and integrated colours for each target (Figure
\ref{sbf_rad}), and comparing the data of the two galaxies between
them and with models, we can make various considerations on the host
unresolved stellar systems. We consider the updated version of the
SPoT models \citep{raimondo11}, which for the photometric bands and
chemical composition used in this section confirm the results obtained
from the previous \citet{raimondo05} models.

As a first general consideration, we find that the SBF
$\bar{V}{-}\bar{R}$ colour, and the integrated $B{-}V$ and $V{-}R$
colours predicted by models representing old ($\gsim$ 7 Gyr) and metal
rich ($[Fe/H]\gsim -0.4$ dex) SSPs are in good agreement with the
measured values. Differently, the $\bar{B}$ is much more sensitive to
the properties of the dominant stellar population (see below) and the
matching with ``canonical'' SSP models is less satisfactory in this
band.

Starting from these general considerations on the properties of the
``dominant'' stellar population in the galaxies (i.e. the stellar
population which is is emitting the largest part for the flux
responsible for the colours and the SBF signal) let us analyse
more in details the result of Figure \ref{sbf_rad}.

The left and middle columns of panels in Figure \ref{sbf_rad} show the
SBF colours and $V{-}R$ measurements in the five annuli considered for
each galaxy versus the integrated $B{-}V$. The annuli appear numbered
in the lower panels according to the numbers reported in Table
\ref{tab_sbf}, so that inspecting the data in the figure it is 
  also possible to recognise the radial behaviour of plotted
quantities. In the third column, instead, the overall SBF measures are
plotted (``main'' annulus data in Table \ref{tab_sbf}, empty symbols)
together with SBF in the SNe regions. It is useful to emphasise
that for both galaxies our integrated colours and colour profiles
agree with the same measurements from \citet{idiart02}.

By inspecting results in the figure, we find that NGC\,4621 does not
show any significant SBF colour gradient either in
$\bar{B}{-}\bar{V}$ or $\bar{V}{-}\bar{R}$ (lower and middle left
panels), and no gradient even in the integrated $V{-}R$ colour (top
left panel). The only colour showing radial changes is the $B{-}V$,
varying from $\sim 0.91$ mag for the inner annulus (\#1), to $\sim
0.84$ in the outermost (\#5). As already mentioned, inspecting
  the $\bar{V}$ and $\bar{R}$ data for NGC\,4621 in Table
  \ref{tab_sbf}, a small but non-negligible gradient seems to be
  present, of $\sim0.25$ mag in $V$ and $\sim0.15$ mag in $R$, if the
  inner annulus is discarded. The $\bar{V}$ and $\bar{R}$ gradients
  have the same trend with colour, i.e. brighter SBF magnitudes
  associated with bluer colours/larger radii, thus they tend to
  cancel-out when the SBF-colour is considered. In fact, if the
  innermost annulus is neglected, a SBF-colour gradient of appears in
  the $\bar{V}{-}\bar{R}$ versus $B{-}V$ plane for NGC\,4621, although
  it is of the same amplitude of estimated uncertainties. Even so, the
  radial trends of the SBF magnitudes and of the $B{-}V$ are both
  consistent with the well known scenario of a more metal-poor stellar
  component (i.e. bluer integrated colours, and brighter SBF
  magnitudes for the pass-bands considered here) at larger
  galactocentric radii.

For NGC\,4374 (central panels) we find that $\bar{V}{-}\bar{R}$ and
$V{-}R$ are practically flat along the galaxy radius, if the innermost
annulus (\#1) is excluded. Due to the presence of a dusty disk in the
galaxy centre, in fact, we left unmasked only a small fraction of the
area in the annulus \#1, but further contamination from dust might
still be present. Differently from the previous case, the $V$ and $R$
data of NGC\,4374 in Table \ref{tab_sbf} do not show any systematicity
versus radius, although the $\bar{V}{-}\bar{R}$ versus $B{-}V$ panel
seems to show a correlation between the two plotted magnitudes, mostly
due to the cited innermost annulus.

As mentioned above, the comparison between measurements and models in
the two upper panels of each column reveals that the data of galaxies
lie close to the position of SSPs with $[Fe/H]\gsim -0.4$ dex and
t$\gsim$7 Gyr. However the age-metallicity degeneracy suffered by
classical and SBF colours obtained by combining these bands prevents
us from giving more precise information on the dominant stellar
  components in the galaxy.

The lower panels, which involve $\bar{B}$, deserve a separate and
detailed discussion, since both galaxies show a $\bar{B}{-}\bar{V}$
colour much bluer than models, an evidence more severe for
NGC\,4621. For this galaxy, in fact, the $(\bar{B}{-}\bar{V})_{Main}$
is $\sim$0.2 mag bluer than NGC\,4374, and $\sim$0.4 mag bluer than
models. We also note that NGC\,4374 shows a radial change leading the
$\bar{B}{-}\bar{V}$ value of the outermost annulus to be quite similar
to the models. Differently, NGC\,4621 does not present any
  systematic gradient and, except for the innermost annulus the
$\bar{B}{-}\bar{V}$ is nearly constant in every studied region of the
galaxy.

To understand the origin of the mismatch between data and models,
visible in the lowermost panels of Figure \ref{sbf_rad}, and of the
peculiarly blue $\bar{B}{-}\bar{V}$ colour of NGC\,4621, we have
considered to take advantage of the versatility of the SPoT stellar
population synthesis code and carry out some specific numerical
simulations, to investigate if and how the presence of a complex
stellar population modifies the expected $\bar{B}{-}\bar{V}$. In
addition, for metal rich models we have also considered the case of a
non negligible fraction of Horizontal Branch (HB) stars having higher
effective temperatures with respect to ``canonical'' stellar evolution
models.

\begin{figure*}
   \centering
   \includegraphics[width=15cm]{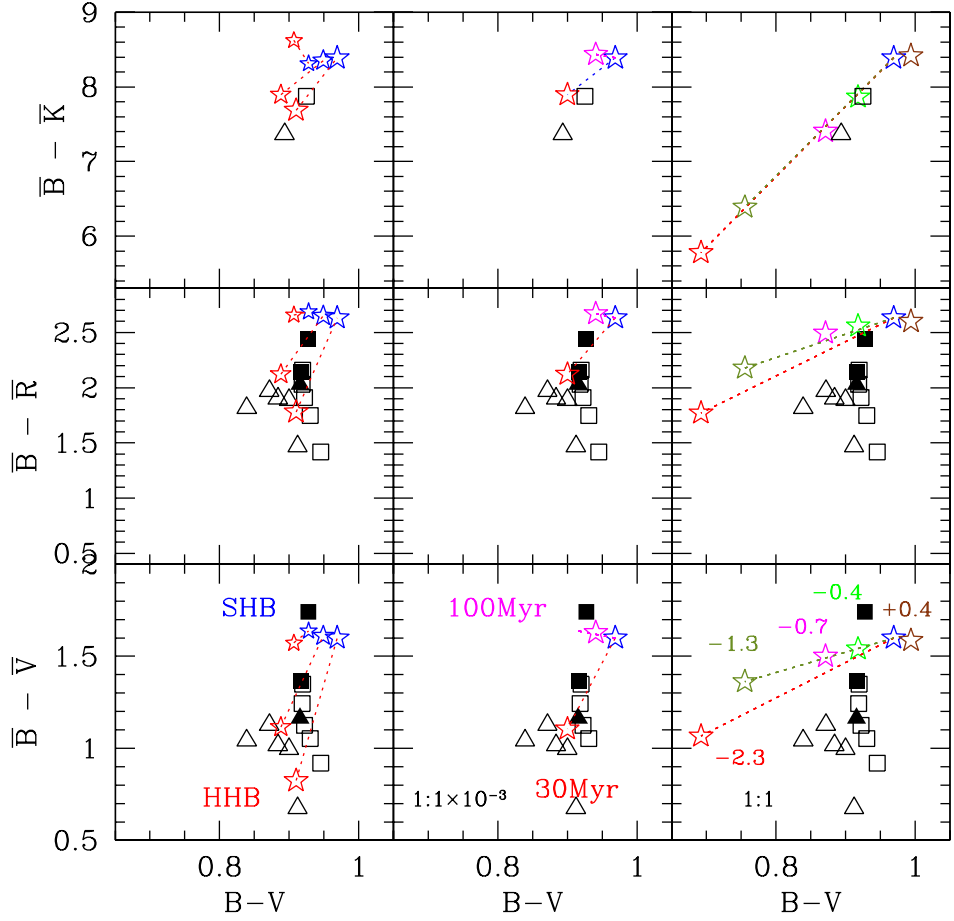}
      \caption{SPoT stellar population models obtained with
        non-canonical assumptions. In all cases the initial (standard)
        population has solar metallicity and t$\sim14$ Gyr, and is
        shown with a blue star in each panel, while the final
        composite population is shown with a different colour and
        connected with a dashed line to the reference model. Upper
        panels show main annulus measures, middle and lower panels
        show multi-annuli measures and SNe\,Ia regions.  Left panels:
        a fraction equal to 50\% of the total HB-stars is simulated
        being HHB. For these models three ages are considered (t=10,
        12 and 14 Gyr, increasing ages are marked with larger
        symbols). Middle panels: a young population of t=30 Myr (red),
        or t=100 Myr (magenta) is added to the old solar one. The
        fraction in mass of old to young stars is reported in the
        lower panel. Right panels: various metallicities are added to
        the solar one, as labelled. The mass fraction metal-poor to
        standard SSPs is shown in the lower panel. Symbols for
        observational data are as in Figure \ref{sbf_rad}.  \it{[See
            electronic version of the Journal for a colour version of
            the figure.]}}
         \label{sbf_hot}
   \end{figure*}

In order to find indications on the origin of the peculiar
$\bar{B}{-}\bar{V}$ behaviour of the two galaxies, our numerical tests
have been organised as follows: $1)$ we have explored the effects of
the presence of a hot HB (HHB) component to field stars; $2)$ a young
stellar population is added to a ``reference'' old and metal-rich
population, and $3)$ a metal-poor population is added to the reference
one. For the comparison in $B$, $V$ and $R$ bands we adopted the SBF
measured in five annuli, plus the measures in the SNa\,Ia regions,
while for the panels including the $\bar{K}$ taken from literature we
adopted our main annuli SBF measurements. The results of the
  simulations and the comparisons with data are briefly discusses
below, and shown in Figure \ref{sbf_hot} where three different SBF
colours are plotted against $B{-}V$.

%_____________________________________________________________
%Figura SBF tests on bar(B-V)
%-------------------------------------------------------------

\begin{itemize}

\item In order to mimic the presence of HHB stars, we carried
  out a numerical experiment by considering that 50\% of HB stars have
  suffered large mass--loss during the RGB phase (first column of
  panels in Figure \ref{sbf_hot}). As a result, a percentage
    of low mass stars populates the blue/hot part of the horizontal
    branch. This is not unrealistic since such hot component is
  observed in several metal-rich stellar clusters (e.g. NGC\,6441 and
  NGC\,6388), whatever the mechanism responsible of this evidence
  is. In the figure we consider three populations with solar
  metallicity and ages 10, 12, and 14 Gyr. Both SBF and integrated
  colours move from the ``standard'' positions (standard HB,
  SHB, label in the figure) to the hot/blue region of the diagram 
    (HHB in the figure). The effect is larger for SBF colours
  including $\bar{B}$ with respect to integrated colour due to their
  increased sensitivity to even a small number of bright hot
  stars.

\item In the second numerical experiment (middle panels) a very young
  population (two different ages are considered: 30\,Myr and 100\,Myr)
  is added to a population of 14 Gyr and solar metallicity, with mass
  ratio between the two components $1_{old}:10^{-3}_{young}$.  From
  the middle panels of Figure \ref{sbf_hot} appears that a very recent
  and {\it diffuse} burst of star formation is required to obtain
    a good matching between models and data.

\item The third column of Figure \ref{sbf_hot} reports the effects of
  a secondary metal-poor component of 14 Gyr added to the main solar
  metallicity one, the $[Fe/H]$ of the metal-poor component are
  labelled in the lower right panel of the Figure. Only a very metal
  poor population with a mass comparable to the one of the main
  component ($1_{m.rich}:1_{m.poor}$) produces relevant effects on SBF
  colours.  In that case, however, the values of the observed
  integrated coluor $B{-}V$ are not well reproduced any more.

\end{itemize}

The three sets of simulations described seem to point out what
already suggested: the most likely solution to the puzzling behaviour
of $B$-band SBF is the presence of hot HB stars. In fact, while the
presence of a hot/young stellar component provides a good matching
between integrated and SBF colours models with data (second scenario,
middle panels in Figure \ref{sbf_hot}), it appears unrealistic that
these regular ellipticals host such diffuse and very young stellar
component. Moreover, in such a case, i.e. an object with massive and
extended recent star formation, it is likely that a large quantity of
dust would still be present, preventing the SBF measure
itself. However, with the only exception of an inner $r\lsim
  20\arcsec$ dust ring in NGC\,4374, we do not find any sign of
extended dust in both galaxies.  The two-metallicity mixing scenario
seems even more unlikely, such mixing, in fact, does not solve
the mismatch between data and models for optical bands.

As a further element in support of the role played by HHB stars
  in determining the SBF signal of NGC\,4621, we recall the works by
  \citet{brown00} and \citet{brown08}, based on deep near and far-$UV$
  images of the compact elliptical galaxy M\,32. Using HST data, the
  authors found that the number of PAGB stars in M\,32 is
  significantly lower than the expectations of their stellar evolution
  models, while the presence of a HHB population has been observed and
  identified as the main contributor to the $UV$-emission of the
  galaxy.
 
Related to this issue, \citet{cantiello07b}, using the $\bar{B}$ data
of eight ellipticals, suggested that the role of hot evolved stars
cannot be neglected in modelling SBF magnitudes in this
pass$-$band. In that case, though, the mismatch between the data and
models in $\bar{B}$ was solved by enhancing the number of PAGB
expected assuming the evolutionary prescriptions by \citet{brocato90}
and \citet{brocato00}.

As far as concerns NGC\,4621, our present results point towards the
direction suggested by the observations of M\,32. We show that HHB
stars can be the stellar component responsible of the observed
$\bar{B}$-excess, even though a combination of two stellar components
(HHB and PAGB stars) could not be ruled out and the topic requires
further investigations.

As already mentioned for the case of M\,32, a further piece of
information comes from the comparison of the integrated properties of
the two galaxies in the wavelength interval where the very hot stellar
component is dominant, i.e. the $UV$ regime. In this wavelength
  regime, \citet{longo91} found that NGC\,4621 is brighter than
  NGC\,4374. The puzzling presence of a strong $UV$ emission in some
regular early-type galaxies, discovered in late $'70s$
\citep{code72,bertola80}, is now widely interpreted as the presence of
a old hot stellar component. Although the mechanisms that created such
component, or its true nature, are not well understood
\citep{park97,kaviraj07,han08}, some of these old hot stellar sources
may have effects on $\bar{B}$, as discussed by various authors based
on both SSP models predictions \citep{worthey93b,cantiello03}, or on
empirical data \citep{shopbell93,sodemann96,cantiello07b}.  Moreover,
a recent study of \citet{buzzoni08} presented a detailed analysis on
the link between the $UV$ characteristics and near--$IR$ SBF
amplitudes of elliptical galaxies.

In conclusion, the present analysis seems to support a scenario where
the peculiar $\bar{B}{-}\bar{V}$ is related to a hot and old diffuse
stellar component, like HHB stars. Larger samples of SBF colours in
blue bands are required to provide further constraints.

\subsection{SBF colour-colour diagrams}

%_____________________________________________________________
%Figura SBF color-colour
%-------------------------------------------------------------
   \begin{figure}
   \centering
   \includegraphics[width=7cm]{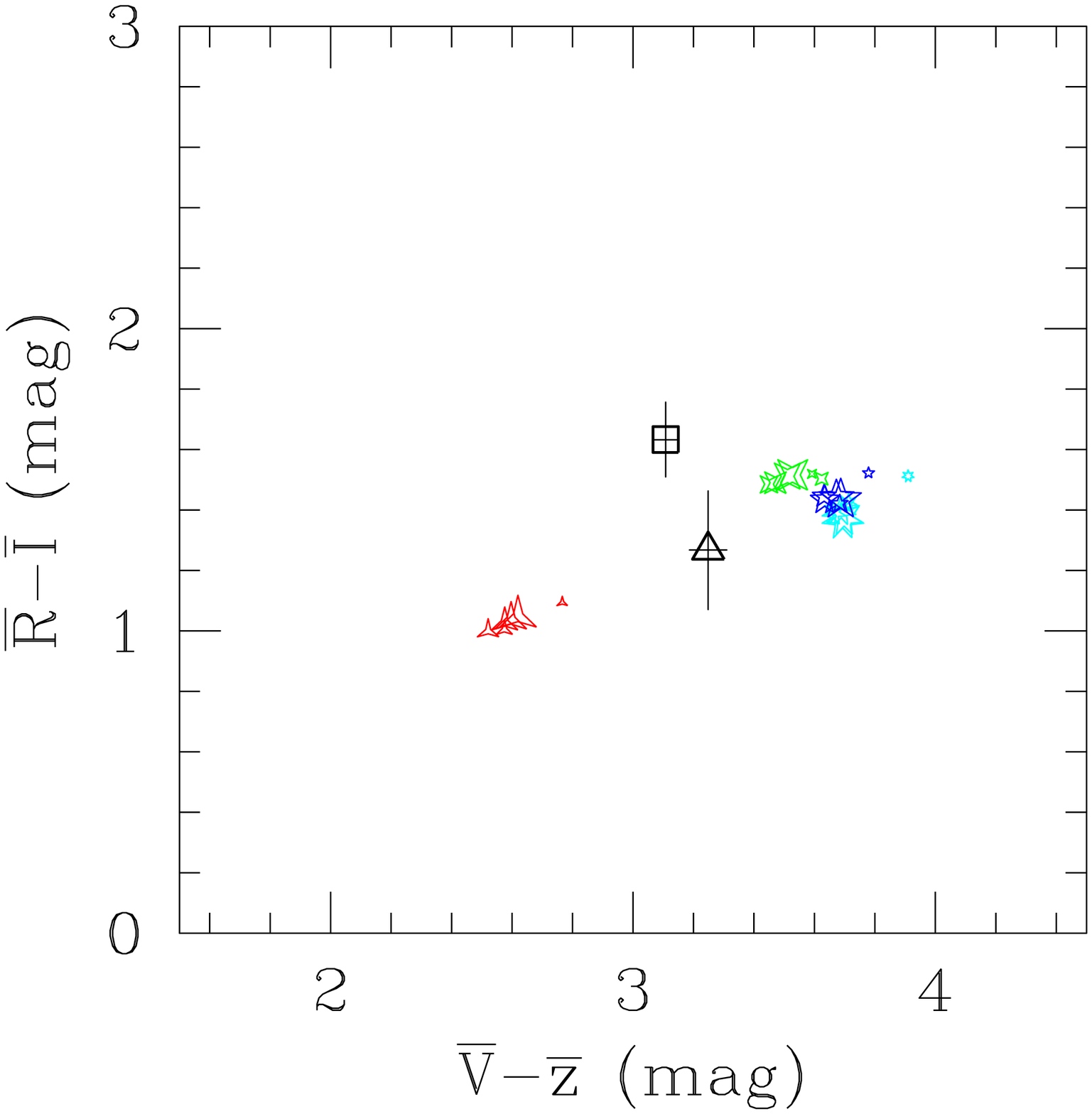} \\
  \includegraphics[width=7cm]{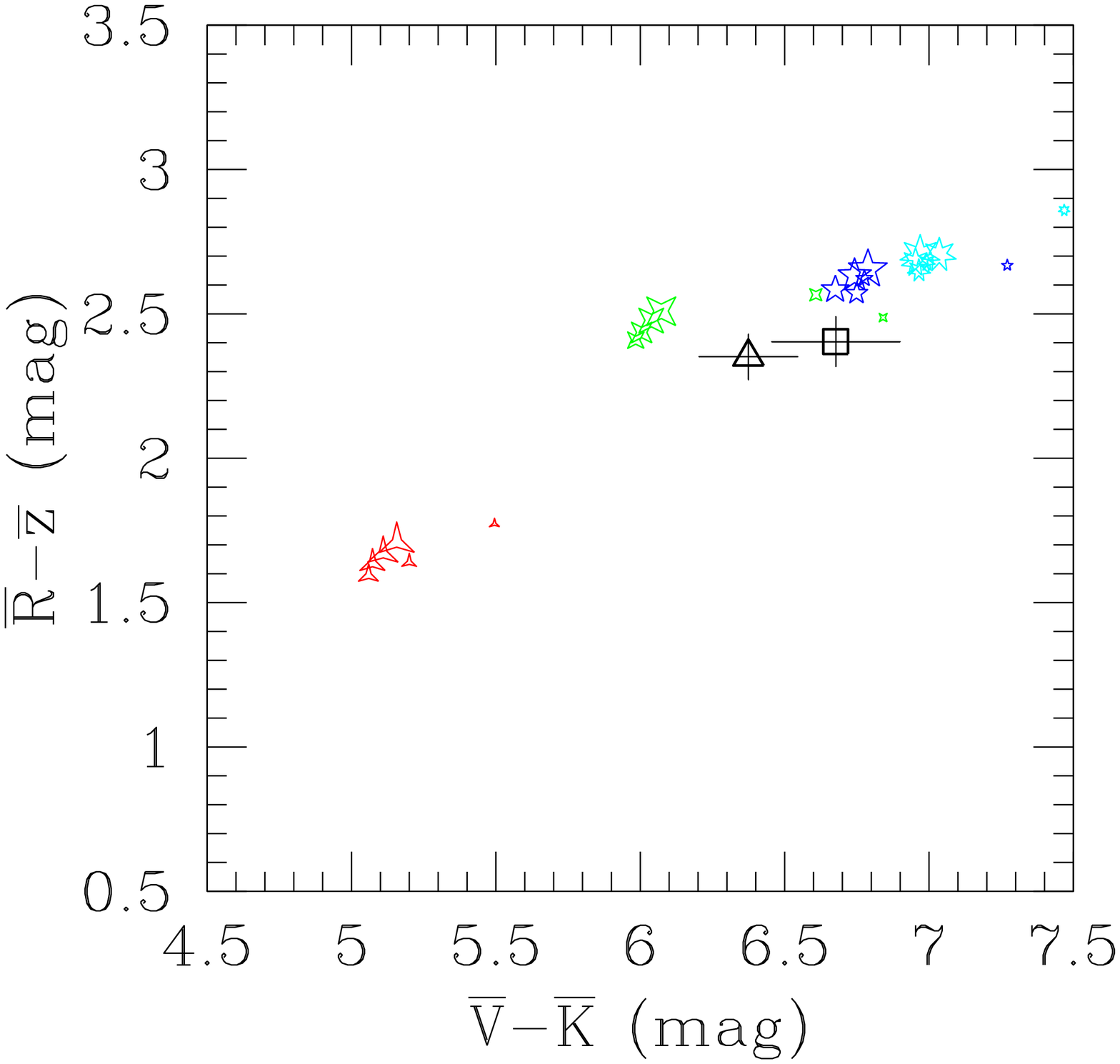} \\
   \includegraphics[width=7cm]{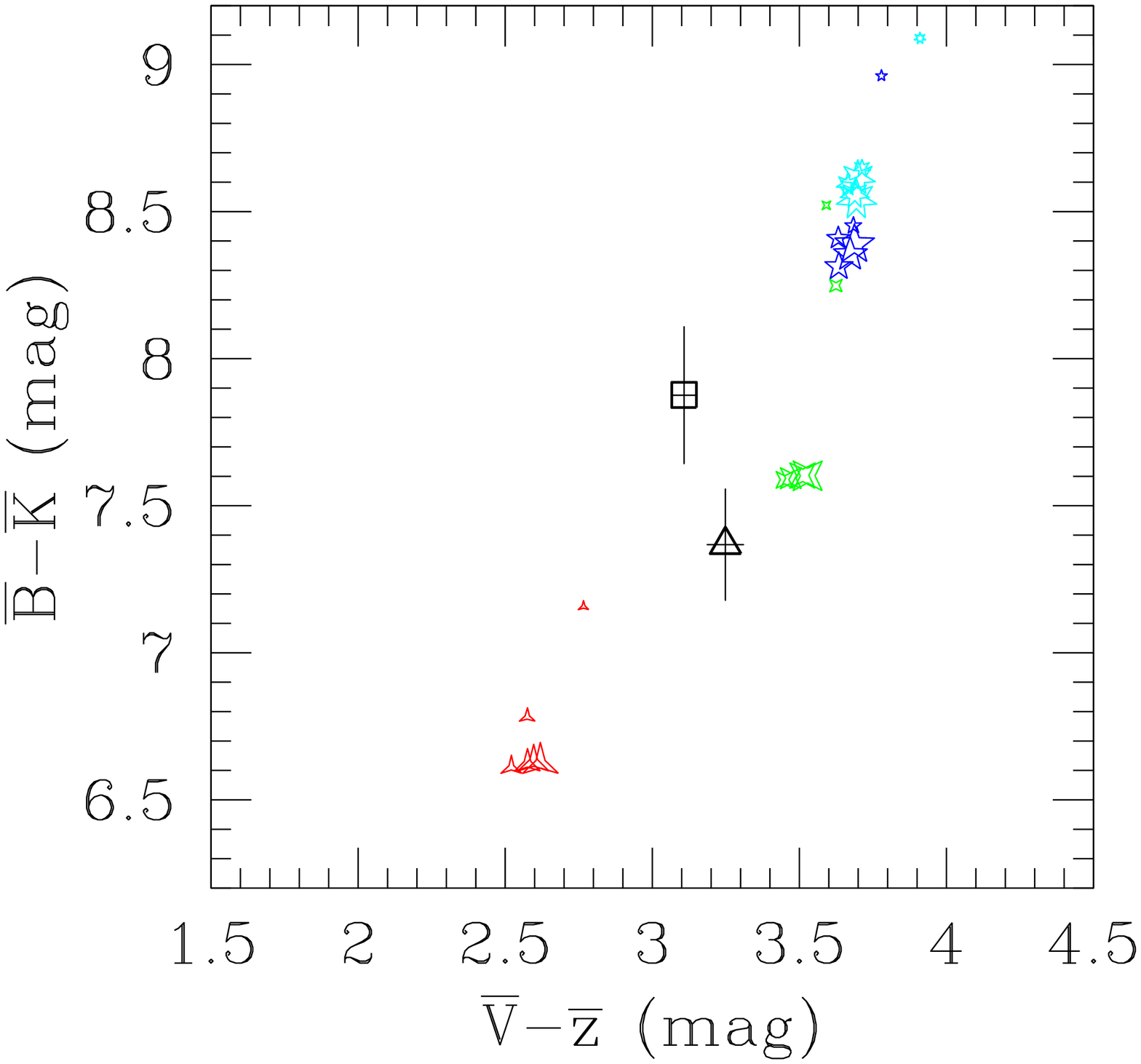} \\
      \caption{SBF colour-colour diagrams obtained from present
      measures (main annulus) and literature data. Symbols and models
      are the same as for Figure \ref{sbf_rad}. For consistency with
      other filters the $z$-band SBF magnitudes given by
      \citet{blake09} are transformed from the ABmag to the Vegamag
      system according to \citet{sirianni05}. \it{[See electronic
      version of the Journal for a colour version of the figure.]}}
         \label{sbf_colors}
   \end{figure}

The coupling of literature $I$, $z$ and $K$-band SBF data for the
  two galaxies with our measures (main annuli) provides a sample of
six independent SBF values useful to obtain three independent
SBF-colour versus SBF-colour diagrams. Such wide pass-band coverage
is, to our knowledge, the first ever presented for SBF. The diagrams
with observations are shown in Figure \ref{sbf_colors}, together with
models.  The upper panel of the Figure shows a feature that has
already been discussed \citep[e.g.][]{bva01,cantiello03}, that is that
model separation for pure optical SBF colours is not effective in
removing the age-metallicity degeneracy. In such panel, in fact, all
$[Fe/H]\geq$-0.4 dex models overlap to each other, with old and metal
rich models superposed to young and metal poor ones.

The models degeneracy is significantly reduced in the middle panel,
which includes the optical to near--$IR$ $V{-}K$ SBF colour. The
position of the two objects overlaps with models with different metal
contents. In particular the bluer point, i.e. NGC\,4621, is
  located near to the region of $[Fe/H]$$\sim$-0.4 dex with respect to
  NGC\,4374, which appears slightly more metal--rich. Such behaviour
is not unexpected, since the latter galaxy is brighter than the first
one and, due to the known mass-metallicity relation
\citep[e.g.][]{bernardi98}, it is reasonable to expect a field
population slightly more metal--rich in NGC\,4374. It is worth to
emphasise that the two-SBF--colour panel represents a useful tool to
provide insights to the ``absolute'' $[Fe/H]$ of the dominant stellar
component, within the observational error-bars, and the SSP models
  scenario adopted.

In other words, based on the SBF-colours analysis and on the
integrated colours of field stars from this work \citep[see
  also][]{tonry01,idiart02,mei07vii} we find that the dominant
stellar population of NGC\,4374 is either more metal rich or older, or
both, than what found in NGC\,4621. Moreover, this result agrees with
independent estimates of the metal content in these galaxies, based on
spectral index measurements, which predict nearly identical
\citep{idiart07} or slightly higher metallicity for NGC\,4621
\citep{kuntschner01,scott09}.

It must be pointed out, though, that these results are based on the
naive interpretation of field stellar population properties solely in
terms of age and chemical content differences. Nevertheless, as
discussed above, SBF magnitudes and colours can be very sensitive to
the presence of a blue/hot stellar sub-system, especially SBF in bluer
pass-bands.

The lower panel in Figure \ref{sbf_colors} includes our $\bar{B}$
measurements. As for the first panel, the $\bar{V}{-}\bar{z}$ colour
is used in abscissa, in this case the separation between different SSP
models allows to recognise that for the two larger $[Fe/H]$ values the
$\bar{V}{-}\bar{z}$ colour is bluer for older ages. Taking as
reference the models with solar metallicity, one can conclude that
both galaxies have similar $\bar{V}{-}\bar{z}$, matching with the
oldest SSP models, but NGC\,4621 has $\bar{B}{-}\bar{K}\sim 0.6$ mag
bluer than NGC\,4374, as one would also expect in the case of a
diffuse hot stellar component is present in this galaxy. As an
example, a canonical SSP model at $t=14$ Gyr, solar metallicity has
$\bar{B}{-}\bar{K}\sim 8.4$ and $\bar{V}{-}\bar{z}\sim3.7$ mag; by
artificially increasing the number of HHB stars (as described in
section \ref{sec_ssp_hot}) our models predict $\bar{B}{-}\bar{K}\sim
7.7$, i.e.  roughly the $\Delta \bar{B}{-}\bar{K}$ colour difference
between the two galaxies, while $\bar{V}{-}\bar{z}$ is left unchanged.

A further consideration is the fact that in the bottom panel of Figure
\ref{sbf_colors}, NGC\,4621 falls within the region of models with
$[Fe/H]$ between -0.7 and -0.4 dex, even though the $B$-band SBF is
too bright. Such behaviour could be explained with bright $K$-band
SBF, so that the $B$ and $K$ SBF excess compensate each other allowing
observations to match with models of intermediate metallicity. This
would also make $\bar{V}{-}\bar{K}$ too red, causing the shift
observed in the middle panel. In that case, i.e. too bright $B$ and
$K$-band SBF, the three panels of Figure \ref{sbf_colors} are all
suggesting a $[Fe/H]\lsim-0.4$ for NGC\,4621, and a larger metallicity
for the dominant stellar component in NGC\,4374.

Whether the position of NGC\,4621 with respect to models is due to the
presence of a metal poor dominant stellar component, or it is related
to a diffuse hot component, cannot be established with present
data. New independent observational data~sets, and accurate modelling,
able to match all the observed properties of the two galaxies, are
needed to better understand the physical characteristics of the
unresolved stellar systems in both targets.

Nevertheless, the coupling of $B$-band with other optical or near--$IR$
SBF measures, appears to be a promising method to unveil the
properties of hot stellar components possibly hidden to other
photometric indicators.

\subsection{SBF colours and stellar populations properties in regions hosting type Ia Supernovae}

The correlation between the SN\,Ia peak magnitude and the host-galaxy
morphology, i.e. the host stellar population, is well known since the
work by \citet{hamuy96}, showing that intrinsically fainter events
occur in early-type galaxies, while luminous events are often seen in
late-type galaxies. The results on the mean stellar population
properties in the galaxies, discussed in the previous sections, seem
to support the idea that old stellar system, i.e. old progenitors, are
required for sub-luminous SNe\,Ia.

To better constrain the properties of the {\it local} stellar
populations we have measured SBF and colours in the areas where the
SNe\,Ia exploded, adopting a $25 \arcsec \times 25 \arcsec$ square
region. The results of the measurements are reported in Table
\ref{tab_sbf}, and shown with full triangles (for NGC\,4621), and full
squares (for NGC\,4374) in the right panels of Figure \ref{sbf_rad}.

First of all, we note that, from $B{-}V$, $V{-}R$ and
$\bar{V}{-}\bar{R}$ data, the stellar population properties in the
selected regions and in the main annulus (empty black symbols) appear
remarkably homogeneous. Taking into account only such colours, we
conclude that the stellar component in the regions hosting the three
SNe\,Ia and the overall stellar population in the galaxies have quite
similar properties, with all data matching SSP models of
$[Fe/H]\gsim-0.4$ dex and $t\gsim7$ Gyr. In any case, old ages are
expected, supporting the results by \citet{gallagher08} based on
spectroscopic data. It is interesting to note that the properties of
the dominant stellar populations in the regions of the two SNe\,Ia
host by NGC\,4374 appear strikingly similar in these panels,
nearly identical to each other, notwithstanding the large projected
separation ($\sim$10 kpc) between the two regions.

Larger differences are seen between the three regions in the
$\bar{B}{-}\bar{V}$ versus $B{-}V$ panel (lower right panel in Figure
\ref{sbf_rad}). Due to the aforementioned stronger sensitivity of
  $B$-band SBF to changes of stellar population properties respect to
  other pass-bands, the larger scatter between the three SNe\,Ia is
  not surprising, and it might be possibly related to different levels
  of local ``pollution'' from a hot stellar component. Furthermore,
  the regions of NGC\,4374 where SN\,1991bg and SN\,1957B exploded
  show a $\Delta \bar{B}{-}\bar{V}\sim 0.4$ mag, despite their nearly
  homogeneous $B{-}V$, $V{-}R$ and $\bar{V}{-}\bar{R}$ colours.
  Whether the $\Delta \bar{B}{-}\bar{V}$ is related to the SNe
  progenitors and their environment, and thus to the $\Delta
  B_{max}\sim 2.3$ mag between the two SNe\,Ia, or to problems with
  the $P_r$ correction cannot be said using present data, and further
analysis based on a richer sample is needed.

For what concerns NGC\,4621, even including $B$-band SBF data we find
a good matching between global and local, i.e. near SN\,1939B, stellar
population properties.

%______________________________________________________________

\section{Conclusions}
\label{sec_end}
We have presented a detailed multi-band study of SBF magnitudes for
two bright galaxies in the Virgo cluster, NGC\,4621 and NGC\,4374,
based on deep $B$, $V$ and $R$-band imaging data of the FORS2
camera at the VLT telescope.

Among the three bands available, the $V$ and $R$ have known empirical
SBF calibrations useful to obtain the distances of the
targets. Coupling our measurements with such calibrations, or with
calibrations obtained from simple and composite stellar
population models, we obtained accurate distances for the two galaxies
which agree very well with other distance estimates taken from
literature. This demonstrates both the reliability of the measured SBF
magnitudes and the goodness of the calibrations adopted.  Taking
advantage of the fact that various SBF distances, based on different
calibrations for different filters, are available for the two targets,
we have compared the average SBF distance moduli with other non-SBF
ones, to check for any possible systematics. Although the sample
of galaxies is statistically small, the results obtained seem to rule
out the presence of possible bias on SBF distances.

Taking the median of all available SBF and non-SBF
distance estimates, we estimated: $m-M=30.98\pm0.17$ for NGC\,4621,
and $m-M=31.22\pm0.16$ for NGC\,4374.

The two target galaxies hosted a total of three SN\,Ia events: one in
NGC\,4621, SN\,1939B, and two in NGC\,4374, SN\,1957B and SN\,1991bg,
all classified as sub-luminous. Using some recent calibration
relations based on decline rate and colours of the SN
\citep{folatelli10}, and on the updated MLCS method \citep{jha07} we
find an excellent agreement between the SBF and SNe\,Ia distances. A
result that is very promising in view of a reduction of the number of
``rungs'' to bridge local to cosmological scale distances, i.e. to
significantly reduce the systematic uncertainties on distances of
objects at large distances and, consequently, on cosmological
parameters.

We also carried out SBF measurements on $B$-band images, but such
measures were rejected for distance analysis. The sensitivity of SBF
magnitudes in this band to the stellar population properties makes
unreliable any calibration in this band. We tentatively derived
a $B$-band calibration based on SPoT models - which have proved being
realistic in both $V$ and $R$ bands, as well as other studied bands
\citep{biscardi08} -, but the results obtained confirm that SBF
magnitudes in such band are not well suited for any kind of distance
analysis.\\

Thanks to the high quality of the data, we have succeeded in measuring
SBF magnitudes in various galactic regions. Both integrated and SBF
colours (excluding $\bar{B}$) seem to point out that the stellar
population are relatively uniform along the galaxy radius. No sizable
SBF gradient is observed in NGC\,4374, while for NGC\,4621 a small but
non-negligible SBF gradient in $V$ and $R$ is observed, consistent
with the known scheme of more metal poor dominant stellar populations
at larger galactocentric radii \citep[e.g.][]{cantiello05}. Given the
fact that $i)$ compared to $I$-band SBF $V$ and $R$ are more sensitive
to stellar population properties \citep{bva01,cantiello03}, and $ii)$
$\bar{I}$ gradients up to $\sim0.4$ magnitudes were measured by
\citet{cantiello05} on similar radial scales for a different set of
targets, the negligible $\bar{V}$ and $\bar{R}$ gradients detected
here might be explained by the position of the two objects within
their host cluster. Since the two targets analysed are located in the
regions with highest galaxy density in the Virgo cluster, in fact,
this behaviour might be related to the major-merging events in such
environment, which tend to smear out possible stellar population
gradients \citep{kobayashi04}.

Taking advantage of the multi-wavelength coverage of our data~set, and
of existing $I$, $z$, and $K$-band SBF measurements, we analysed the
data of both galaxies in the SBF--colour versus integrated--colour and
SBF--colour versus SBF--colour diagrams. This is the first
SBF-analysis carried out using SBF data in six different pass-bands.
As a result we concluded that the dominant stellar component in the
two galaxies is very similar, though NGC\,4374 seems to be slightly
more metal rich than NGC\,4621.

If $B$-band measures are taken into account, the SBF colours of the
two galaxies show non-negligible differences, with NGC\,4621 having
brighter SBF than NGC\,4374.  Given the known link between SBF
magnitudes in blue bands and the properties of a hot old stellar
component \citep[][]{worthey93b,cantiello03}, we used the SPoT Stellar
Population Synthesis code to simulate populations with
``non-canonical'' properties. In particular, to a solar metallicity
old $t\sim14$ Gyr population, we have $1)$ enhanced the content of hot
HB stars (i.e., $\sim$ 50\% canonical HB and $\sim$ 50\% HHB), $2)$
added a very young (down to 30 Myr) diffuse secondary component, and
$3)$ added a more metal poor SSP (down to $[Fe/H]\sim-2.3$
dex). Within the limits of the small number of data, and adopting as a
razor the physical plausibility of the population mixing, the
simulations seem to favour the HHB component scenario.

This is also supported by the fact that \citet{brown00} and
\citet{brown08} find the presence of HHB stars in the compact
elliptical galaxy M\,32 on the basis of $UV$ observations. In this
framework, it is relevant to recall that NGC\,4621 is substantially
brighter than NGC\,4374 in the $UV$ bands \citep{longo91}, and that
the integrated properties of normal elliptical galaxies with bright
$UV$ emission are now interpreted as an effect of a diffuse hot
stellar component \citep[e.g.][]{park97}. Thus, further analysis is
required to better constrain the links between the $UV$ emission and
SBF in early type galaxies, also in view of the recent studies on the
relation of $UV$ with near--$IR$ SBF amplitudes \citep{buzzoni08}, and
larger samples of SBF colours in blue bands are required to provide
further constraints to this scenario.

Finally, taking into account the integrated colours and $\bar{V}$ plus
$\bar{R}$ SBF for the areas of the galaxies where type Ia Supernova
exploded, we find no substantial differences between the local and the
global stellar population properties in the galaxies. For SN\,1939B
and its host, NGC\,4621, this is also true for $B$-band SBF data. On
the contrary, there is a $\Delta \bar{B}{-}\bar{V}\sim$0.4 mag
difference between the regions of the two SNe host in NGC\,4374,
SN\,1957B and SN\,1991bg. Due to the quoted sensitivity of blue-band
SBF to stellar population properties, such difference and, more in
general, the scatter in $\bar{B}{-}\bar{V}$ between the three SNe\,Ia
might be related to different levels of local pollution from a hot
stellar component, although for the case of SN\,1991bg we cannot
exclude the possible systematic effect of a low $\bar{B}$ signal,
comparable to the $P_r$ correction.

In conclusion, the present set of SBF measures shows that optical to
near--$IR$ SBF magnitudes can be very effective to unveil the properties
of global and/or local stellar populations in distant galaxies. In
particular, if $\bar{B}$ (or even SBF in bluer bands) are available,
various constraints can be set to the role of hot field stars in
normal ellipticals, at the same time multi--band SBF can be used to
analyse the relation between SNe and local stellar population
properties. Nevertheless, serious limitations are set by the small
samples of measures still available, new observations are advisable to
further study stellar population properties based on multi--colour SBF
analysis.

\begin{acknowledgements}
M.C. acknowledges the support provided by the PRIN-INAF 2008
(PI. M. Marconi).  We also thank the referee for his very helpful
  suggestions. Based on data obtained from the ESO Science Archive
  Facility.

\end{acknowledgements}

%%%%%%%%%%%%%%%%%%%%%%%%%%%%%%%%%%%%%%%%%%%%%%%%%%%%%%%
%%%%%%%%%%%BIBLIOGRPHY %%%%%%%%%%%%%%%%%%%%%%%%%%%%%%%%
%%%%%%%%%%%%%%%%%%%%%%%%%%%%%%%%%%%%%%%%%%%%%%%%%%%%%%%

\bibliography{cantiello_may10}
\bibliographystyle{aa}

\end{document}